%% file: main.tex
\documentclass[acmsmall]{acmart}

\AtBeginDocument{%
  }

\setcopyright{rightsretained}
\acmJournal{PACMSE}
\acmYear{2025} 
\acmVolume{2}
\acmNumber{FSE} 
\acmArticle{FSE032} 
\acmMonth{7} 
\acmDOI{10.1145/3715749}

\input{tool}
\begin{document}

\title{Beyond Functional Correctness: Investigating Coding Style Inconsistencies in Large Language Models}

\newcommand\corrauthorfootnote[1]{%
  \begingroup
  \renewcommand\thefootnote{}\footnote{\textsuperscript{*}#1}%
  \addtocounter{footnote}{-1}%
  \endgroup
}


\author{Yanlin Wang}
\orcid{0000-0001-7761-7269}
\affiliation{%
  \institution{Sun Yat-sen University}
  \city{Zhuhai}
  \country{China}
  }
\email{wangylin36@mail.sysu.edu.cn}

\author{Tianyue Jiang}
\orcid{0009-0003-1305-8084}
\affiliation{%
  \institution{Sun Yat-sen University}
  \city{Zhuhai}
  \country{China}
  }
\email{jiangty9@mail2.sysu.edu.cn}

\author{MingWei Liu}\authornote{Corresponding Author}
\orcid{0000-0002-3462-997X}
\affiliation{%
  \institution{Sun Yat-sen University}
  \city{Zhuhai}
  \country{China}
  }
\email{liumw26@mail.sysu.edu.cn}

\author{JiaChi Chen}
\orcid{0000-0002-0192-9992}
\affiliation{%
  \institution{Sun Yat-sen University}
  \city{Zhuhai}
  \country{China}
  }
\email{liumw26@mail.sysu.edu.cn}

\author{MingZhi Mao}
\orcid{xxx}
\affiliation{%
  \institution{Sun Yat-sen University}
  \city{Zhuhai}
  \country{China}
  }
\email{mcsmmz@mail.sysu.edu.cn}

\author{Xilin Liu}
\orcid{0009-0001-4870-1012}
\affiliation{%
  \institution{Huawei Cloud Computing Technologies}
  \city{Shenzhen}
  \country{China}
  }
\email{liuxilin3@huawei.com}

\author{Yuchi Ma}
\orcid{0009-0002-3304-1389}
\affiliation{%
  \institution{Huawei Cloud Computing Technologies}
  \city{Shenzhen}
  \country{China}}
\email{mayuchi1@huawei.com}

\author{Zibin Zheng}
\orcid{0000-0002-7878-4330}
\affiliation{%
  \institution{Sun Yat-sen University}
  \city{Zhuhai}
  \country{China}}
\email{zhzibin@mail.sysu.edu.cn}

\renewcommand{\shortauthors}{Wang et al.}

\begin{abstract}
Large language models (LLMs) have brought a paradigm shift to the field of code generation, offering the potential to enhance the software development process. However, previous research mainly focuses on the accuracy of code generation, while coding style differences between LLMs and human developers remain under-explored. In this paper, we empirically analyze the differences in coding style between the code generated by mainstream LLMs and the code written by human developers, and summarize coding style inconsistency taxonomy. 
% We adopt the previous work's definition of coding style, but provided a more detailed and refined classification of coding style. 
Specifically, we first summarize the types of coding style inconsistencies by manually analyzing a large number of generation results. We then compare the code generated by LLMs with the code written by human programmers in terms of readability, conciseness, and robustness. The results reveal that LLMs and developers exhibit differences in coding style. Additionally, we study the possible causes of these inconsistencies and provide some solutions to alleviate the problem. 
\end{abstract}





\maketitle

\input{body}

\bibliographystyle{ACM-Reference-Format}
\bibliography{ref}

\end{document}

%% file: tool.tex
\usepackage{float}
\usepackage{array}
\usepackage{rotating}  
\usepackage{pdflscape}
\usepackage{tabularx, multirow, makecell}

\usepackage{adjustbox}
\usepackage{amsmath,amssymb,amsfonts}
\usepackage{algorithmic}
\usepackage[linesnumbered,ruled,vlined]{algorithm2e}
\usepackage{setspace}
\usepackage{graphicx}
\usepackage{textcomp}
\usepackage{xcolor}
\usepackage{multirow}
\usepackage{balance}
\usepackage[framemethod=TikZ]{mdframed}
\usepackage{xcolor}
\usepackage{booktabs}
\usepackage{tcolorbox}
\usepackage{threeparttable}
\usepackage{colortbl}
\usepackage{subfigure}
\usepackage{hhline}
\usepackage{pifont}
\usepackage{listings}
\usepackage{enumitem}
\usepackage{geometry}
\usepackage{verbatim}
\usepackage{xspace}
\usepackage{listings}
\newcolumntype{C}{>{\centering\arraybackslash}X}
\usepackage{pgfplots}
\pgfplotsset{compat=1.16}
\usetikzlibrary{pgfplots.statistics}

\definecolor{ggray}{HTML}{eff0f0}
\definecolor{gggray}{HTML}{E8E8E8}
\definecolor{ggggray}{HTML}{BEBEBE}

\newcommand{\gptfour}{GPT-4\xspace}

\newcommand{\deepsmall}{DeepSeekCoder-1.3B\xspace}
\newcommand{\deepbig}{DeepSeekCoder-6.7B\xspace}
\newcommand{\codellama}{CodeLlama-7B\xspace}
\newcommand{\starcoder}{StarCoder2-7B\xspace}

\newcommand{\NamingFormat}{Naming Format\xspace}
\newcommand{\NamingSemantics}{Naming Semantics\xspace}
\newcommand{\Space}{Space\xspace}

\definecolor{myyellow}{HTML}{FFF2CC}

\newcounter{finding}

\newcommand{\boxmargin}{3.5mm}
\tcbuselibrary{most, skins, breakable, theorems}
\newtcolorbox{myboxa}[2][]{
    colback=gray!10!white,
    colframe=black, enhanced,
    attach boxed title to top left={yshift=-2mm,xshift=5mm},
    title=#2,#1
}
\newtcolorbox{myboxb}[2][]{
    boxsep=3pt,
    left = \boxmargin, right = \boxmargin, top = \boxmargin, bottom = \boxmargin,
    title={#2},#1
}
\newtcolorbox{myboxc}{
    colback=gray!15!white,
    arc = 0pt, outer arc = 0pt,
    boxsep=0pt, left = 3pt, right = 0pt, top = 0pt, bottom = 0pt, 
    leftrule=3pt, bottomrule=0pt,toprule=0pt, rightrule=0pt,
    left = \boxmargin, right = \boxmargin, top = \boxmargin, bottom = \boxmargin
}
\newtcolorbox{myboxd}{
    colback=gray!10,%gray background
    colframe=black,% black frame colour
    width=\columnwidth,% Use 8cm total width,
    arc=1mm, auto outer arc,
    boxrule=0.5pt,
}
\usepackage{tikz}
\newcommand*\blackcircled[1]{\tikz[baseline=(char.base)]{
            \node[shape=circle,draw,inner sep=0.8pt,fill=black,text=white] (char) {#1};}}

\newcommand{\figmargin}{\vspace{1pt}}

%% file: body.tex
\section{Introduction}
% task introduction/importance
Code generation is to automatically generate code snippets that align with given requirements, which plays a vital role in the software engineering domain~\cite{dong2023self,zan2023large,olausson2023demystifying,le2022coderl,zan2023private,huang2023enhancing,jiang2021exploring, li2023think, bairi2023codeplan,chen2023improving,li2023structured,yadav2023exploring,li2023large,li2023towards,jiang2023self,zhang2023self, shi2023towards,mu2023clarifygpt,survey1,jiang2023selfevolve,zhang2023planning}.
Recently, the advent of large language models ~\cite{nijkamp2023codegen,li2023starcoder,guo2024deepseekcoder}, such as CodeLlama~\cite{roziere2023code}, StarCoder~\cite{li2023starcoder}, Codex~\cite{openai2021codex} and \gptfour{}~\cite{achiam2023gpt}, has greatly advanced the performance of code generation. These models have demonstrated remarkable capabilities in code generation, significantly enhancing software development efficiency \cite{zheng2023survey}. However, while previous studies primarily focus on enhancing the accuracy of LLM-based code generation, an equally important aspect—coding style—remains under-explored. Understanding the differences in coding style between LLMs and human developers is essential, as it directly impacts code readability, maintainability, and overall software quality \cite{mi2016measuring}. 

There are several previous works related to coding style~\cite{oman1990taxonomy,parr2016towards,markovtsev2019style,mi2016measuring,chen2023duetcs}.
Oman et al.~\cite{oman1990taxonomy} proposed a programming style taxonomy, including typographic style, control structure style, and information structure style. However, some of these guidelines are outdated, such as the use of Goto statements, which are now rarely used in modern programming.
To improve code formatting style, 
tools like CODEBUFF~\cite{parr2016towards}, an automatic code formatter, and STYLE-ANALYZER~\cite{markovtsev2019style}, which fixes formatting inconsistencies, have been developed. 
Mi et al.~\cite{mi2016measuring} expanded the scope by using hierarchical agglomerative clustering to measure stylistic inconsistency, considering not only formatting but also stylistic metrics related to code readability and features specific to the C/C++ programming languages. More recently, DUETCS~\cite{chen2023duetcs} was proposed for coding style transfer. This work considers a broader range of coding style features, categorizing them into text style (formatting and naming conventions) and structure style (code blocks ordering and preferences of control flow statements). These works provide a preliminary foundation and inspiration for examining coding styles.
\textbf{In our work, we adopt the definition of code style in previous research~\cite{mi2016measuring, berry1985style, yasir2022exploring, oman1990taxonomy}: Code style can be understood as a personal habit in writing source code, reflecting individual preferences in aspects like physical layout, algorithms, etc.} However, there remain several gaps. \textbf{(i)} Existing work offers a generalized definition of code style but lacks detailed and comprehensive classification. \textbf{(ii)} No existing studies have yet analyzed the differences in coding style between mainstreams LLMs and human developers. \textbf{(iii)} A comparative analysis of coding styles across different LLMs is still missing.

% our work.
In this paper, our aim is to fill these gaps.
\textcircled{1} Firstly, we conduct extensive manual analysis to categorize various types of coding style inconsistencies. We compare the code generation results of five mainstream LLMs, including four Code LLMs and a state-of-the-art general-purpose LLM (\gptfour{}~\cite{achiam2023gpt}), against the ground truths of CoderEval~\cite{yu2024codereval} benchmark.
We annotate the results and perform open coding to develop a comprehensive taxonomy of coding style inconsistencies, achieving a detailed classification of code style, which addresses \textbf{(i)} and \textbf{(ii)}. \footnote{In this paper, we use ``coding style inconsistency'', ``style inconsistency'', and ``inconsistency'' interchangeably.}
\textcircled{2} Secondly, we analyze the distribution of inconsistencies, including ratios, frequencies, and differences across the five LLMs, addressing \textbf{(iii)}.
\textcircled{3} Thirdly, we compare the generated code with human-written code based on readability, conciseness, and robustness. 
\textcircled{4} Finally, we experiment on several prompting strategies to explore methods to improve the coding style of LLMs. 

% evaluation results
Through extensive experiments and evaluation, we have obtained the following results:
\blackcircled{1} We propose the first taxonomy of coding style inconsistencies in LLM-generated code. The taxonomy contains 24 inconsistency types, categorized into five dimensions, i.e., \textit{Formatting Inconsistency}, \textit{Semantic Inconsistency}, \textit{Expression/Statement Inconsistency}, \textit{Control Follow Inconsistency}, and \textit{Fault Tolerance Inconsistency}.
\blackcircled{2} Our analysis reveals significant style inconsistencies between human-written code and all studied LLMs, particularly in statements/expressions and formatting. Coding styles among the LLMs themselves are generally similar, with some variation in formatting.
\blackcircled{3} Overall, LLM-generated code is comparable to human-written code in terms of readability, conciseness, and robustness. 
\blackcircled{4} Certain prompt strategies can marginally improve the readability and robustness of generated code, but there is a trade-off with conciseness. This suggests that while prompt engineering can help, it may not fully resolve coding style issues and can sometimes reduce code accuracy.
\blackcircled{5} Our case studies show that LLM-generated code often exhibits poor formatting, lacks familiarity with basic Python functions, and underutilizes advanced syntax features, making the code less readable, concise, and efficient.

% contributions
We summarize the main contributions of this paper as follows:
\begin{itemize}%[leftmargin=8pt]
\item We provide a comprehensive taxonomy of coding style inconsistencies between LLMs and human developers. Compared to previous work, our proposed taxonomy achieves a more detailed classification of coding styles.
\item We conduct extensive analysis to reveal the coding style inconsistencies of several mainstream LLMs, leading to a deeper understanding of LLM-based code generation.
\item We propose some practical solutions to improve the inconsistencies of the coding style, paving the way for a more harmonious integration of LLM and coding practices.
% \item We provide the code and data at \url{https://anonymous.4open.science/r/LLMCodingStyle/}.
\end{itemize}

\section{Related Work}
\subsection{LLM-based Code Generation}

Code LLMs, such as StarCoder~\cite{li2023starcoder}, CodeLlama~\cite{roziere2023code}, and DeepSeek-Coder~\cite{guo2024deepseekcoder}, are specifically optimized for code-centric tasks~\cite{zheng2023survey, zheng2023towards}, leveraging massive code-specific corpora and specialized training instructions. The most state-of-the-art general-purpose large model, GPT-4, due to its significantly larger number of parameters compared to other Code LLMs, demonstrates outstanding performance on code-centric tasks. 
In recent years, some works have studied the application of LLMs in fields such as vulnerability detection~\cite{cheshkov2023evaluation, wang2024m2cvd, tamberg2024harnessinglargelanguagemodels, yang2024securityvulnerabilitydetectionmultitask, yusuf2024instructionshelpfulassessingefficacy}, commit message generation~\cite{lopes2024commit, mandli2024comet, tao2024kadel, zhang2024automatic}, unit test generation~\cite{siddiq2023exploring, xie2023chatunitest, yuan2023no, schafer2023empirical}, code search~\cite{kondo2024improving, wang2023you, hu2024tackling, li2024procqa, gong2024cosqa+}, code summarization~\cite{sun2023automatic,ahmed2024automatic, su2024distilled, li2024machines, haldar2024analyzing, wang2024sparsecoder} and code generation~\cite{DBLP:conf/kbse/LiuYLDWP23, yuan2023evaluating, liu2024stall+, huang2024karecoder, zhu2024hot, tipirneni2024structcoder, ugare2024improving, sun2023don, jain2023llm, ni2023lever, li2023skcoder, wang2023chatcoder, guo2024stop, wang2024rlcoder, li2024repomincoder}, etc.

To understand the code generation performance of LLMs, some high-quality code generation benchmarks have been proposed in recent years. For example, HumanEval~\cite{chen2021evaluating}, MBPP~\cite{austin2021program} and ClassEval~\cite{du2023classeval}, covering different scenarios such as repository-level code generation~\cite{zhang2023repocoder, phan2024repohyper, li2024deveval, li2024evocodebench} and class-level code generation tasks~\cite{du2023classeval}. 
While most studies are primarily concerned with improving the functional correctness of code generated by models, using metrics like pass\@k~\cite{chen2021evaluating}, recent research has begun to explore other attributes of code generated by LLMs. 
For instance, methods have been proposed to enhance the robustness of LLMs~\cite{chen2022generating, zhang2022towards}, and attention has been given to the security aspects of LLMs in code generation tasks, investigating potential vulnerabilities and risks~\cite{oh2023poisone, co2023vulnerabilities}.

In contrast to previous works, our investigation is on the code style of LLMs. We conduct the first study to compare the code style of several mainstream  LLMs with code written by human programmers. 
Additionally, we compare the code styles among different mainstream LLMs.
This analysis provides insights into the strengths and weaknesses of LLMs in terms of coding style, shedding light on potential areas for improvement and future research directions.

\subsection{Coding Style}

% \todo{(1) Add our definition of code style in the following paragraph}

In previous work~\cite{oman1990taxonomy}, Oman et al. proposed a programming style taxonomy, encompassing typographic style, control structure style, and information structure style, which laid the foundation for the development of programming style guidelines and analyzers. However, certain rules in this taxonomy may now be considered outdated and may not fully reflect modern programming practices and conventions (For instance, the example rules mentioning the use of Goto statements are no longer widely used in contemporary programming). Recent strides in coding style research include innovations like CODEBUFF~\cite{parr2016towards}, an automatic code formatter that leverages machine learning to understand and apply code formatting styles.
Similarly, STYLE-ANALYZER~\cite{markovtsev2019style} addresses code formatting inconsistencies using a decision tree forest model. However, both CODEBUFF and  STYLE-ANALYZER focus solely on formatting style. Mi et al.~\cite{mi2016measuring} employed hierarchical agglomerative clustering to gauge code style inconsistencies, focusing on C/C++ languages. In a recent study~\cite{chen2023duetcs}, DUETCS extracted comprehensive code style features from target code examples, covering text and structure style elements. DUETCS utilizes a Siamese feature network to transform source code style into that of target examples while preserving semantic integrity.

% Previous work only provides a generalized definition of code style, but the classification of code style remains insufficiently detailed and comprehensive. 
% In our work, we adopt the definition of code style from some previous studies~\cite{mi2016measuring, yasir2022exploring}: 

% Unlike previous studies, our work represents the first empirical examination of coding style inconsistencies between code generated by Code LLMs and code written by human programmers. Drawing on established coding style categories and definitions from prior literature, we conducted open coding on samples generated by several mainstream Code LLMs. This process yielded a coding style inconsistency taxonomy comprising five dimensions and 24 distinct inconsistency types. In comparison to prior efforts, our proposed taxonomy of code style inconsistencies is more comprehensive and detailed, extending beyond traditional considerations of text style and structure. Furthermore, our study lays the groundwork for future research on the coding style of Code LLMs, offering valuable insights and avenues for further exploration in this field.

In our study, we adopt the definition of code style provided by earlier works~\cite{mi2016measuring, berry1985style, yasir2022exploring, oman1990taxonomy}: Code style can be understood as a personal writing habit in source code, reflecting individual preferences in aspects like physical layout, algorithms, and more. While previous research offers a generalized definition of code style, the classification of code style lacks sufficient detail and comprehensiveness. To address this gap, we conducted open coding on code samples generated by several mainstream LLMs. This process result a coding style inconsistency taxonomy comprising 5 dimensions and 24 distinct inconsistency types, offering a comprehensive classification for code style. In comparison to prior efforts, our proposed taxonomy of code style inconsistencies is more comprehensive and detailed, extending beyond traditional focus on text style and structure style. Moreover, our study lays the groundwork for future research on the coding style of LLMs, offering valuable insights and avenues for further exploration in this domain.

% In our study, we adopt the definition of code style provided by earlier works~\cite{mi2016measuring, yasir2022exploring}: Code style refers to the personal habits reflected in source code, showcasing individual preferences in areas such as physical layout, algorithms, and more. While previous research provides a generalized definition of code style, it lacks a sufficiently detailed and comprehensive classification. To address this gap, we applied open coding to code samples generated by several mainstream large language models (LLMs). This process resulted in a taxonomy of coding style inconsistencies, comprising five dimensions and 24 distinct types. Our taxonomy offers a more thorough and nuanced classification of code style compared to previous efforts, extending beyond the traditional focus on text style and structure. Moreover, our study provides a foundation for future research into the coding style of code-oriented LLMs, offering valuable insights and potential avenues for further exploration in this domain.

\section{Experimental Setup}
\label{sec:setting}
In this section, we introduce the experimental setup, including the LLM selection, benchmark selection, and implementation details.

\subsection{LLM Selection}
\label{sec:setting:models}

% \begin{mdframed}[backgroundcolor=yellow!25,hidealllines=true,innerleftmargin=0pt,innerrightmargin=0pt,innertopmargin=0pt,innerbottommargin=0pt, skipabove=0pt, skipbelow=0pt]
% (1) Add gpt-4 related content
% \end{mdframed}

% We select four mainstream and representative open-sourced Code LLMs 
% that have demonstrated strong performance in the code generation task, namely \codellama{}, \starcoder{}, \deepsmall{}, and \deepbig{}. Due to the constraints in computing resources, we exclude larger models with more than 7 billion parameters. 
% The models we selected are all base models without instruction-tuning, which is particularly suitable for our code completion scenario, wherein the task is to complete the code based on the given context.
% For the four selected Code LLMs, we directly obtain and run their released versions from their official repositories, following the provided documentation. The same settings are being used for all LLMs. 

We select five  LLMs for our experiments, including four well-known Code LLMs (\codellama{}~\cite{roziere2023code}, \starcoder{}~\cite{li2023starcoder}, \deepsmall{}~\cite{guo2024deepseek}, and \deepbig{}~\cite{guo2024deepseek}) and the state-of-the-art general-purpose model, \gptfour{}~\cite{achiam2023gpt}. The rationale behind choosing these four Code LLMs is their demonstrated strong performance in code generation tasks. As for \gptfour{}, it is chosen due to its status as the most effective and parameter-intensive model.
The four selected code LLMs are base models without instruction-tuning, making them particularly well-suited for our code generation scenario, where the objective is to generate code based on the provided context.

% For the selected LLMs, we directly obtained and executed the released versions from their official repositories, following the corresponding documentation. 
% Consistent settings were applied across all five models for a fair comparison.

% \begin{mdframed}[backgroundcolor=yellow!25,hidealllines=true,innerleftmargin=0pt,innerrightmargin=0pt,innertopmargin=0pt,innerbottommargin=0pt, skipabove=0pt, skipbelow=0pt]
% (1) Add content: It should be noted in the article that the projects to which our selected tasks belong do not have any special requirements on code style, so the labeling results are negligibly affected by the requirements of each project on code style.
% \end{mdframed}

\subsection{Benchmark Selection}
\label{sec:setting:tasks}
Our experiments are conducted on \textbf{CoderEval}~\cite{yu2024codereval}, which is a benchmark used to evaluate code generation performance on pragmatic code generation tasks, i.e., code generation with repository context. It consists of 230 Python and 230 Java tasks from real-world open-source projects. Each task contains a function signature\footnote{We use ``method'' and ``function'' interchangeably in this paper.}, a task description, a solution code as the ground truth, and several unit tests to assess the functional correctness of the generated code. 
\textbf{It is worth noting that the repositories to which these Python tasks belong do not have specific requirements for code style.} The objective of each task is to complete the code specified by the function signature, guided by the provided task description, and ensure that it passes the associated unit tests. In this study, we focus on Python tasks due to Python's popularity~\cite{srinath2017python} and its alignment with previous code generation work~\cite{bafatakis2019python}. 

% \subsection{Experimental Details}
% \begin{mdframed}[backgroundcolor=yellow!25,hidealllines=true,innerleftmargin=0pt,innerrightmargin=0pt,innertopmargin=0pt,innerbottommargin=0pt, skipabove=0pt, skipbelow=0pt]
% (1) Add gpt-4 related content
% \end{mdframed}
\subsection{Implementation Details}
% During the inference stage, we configure the five LLMs with identical hyperparameter settings. We employ a random sampling strategy, setting the maximum context length to 1024 and the temperature to 0.8. Each model generates one output per inference. The choices for maximum context length and temperature are aligned with the experimental setup used in CoderEval~\cite{yu2024codereval}. All experiments are conducted on a machine with 216 GB of RAM and a Tesla A100 GPU with 80 GB of memory.

During the inference stage, for the four Code LLMs, we employ a random sampling strategy, setting the maximum context length to 1,024 and the temperature to 0.8. For GPT-4, we adopted identical hyperparameter settings. The choices for maximum context length and temperature are aligned with the experimental setup used in CoderEval~\cite{yu2024codereval}. All experiments are conducted on a machine with 216 GB of RAM and a Tesla A100 GPU with 80 GB of memory.

\section{Evaluation}
In this section, we report and analyze the experimental results to answer the following research questions (RQs):
\begin{itemize}%[leftmargin=10pt]
    \item \textbf{RQ1:} What are the types of coding style inconsistencies between the selected LLMs and human? 
    \item \textbf{RQ2:} What is the distribution of the coding style inconsistencies?
    \begin{itemize}%[leftmargin=15pt]
        \item \textbf{RQ2.a}: What are the percentages of inconsistent coding style for different models?
        \item \textbf{RQ2.b}: What are the inconsistency type numbers present between a single code sample and the corresponding ground truth?
        \item \textbf{RQ2.c}: What are the distribution of coding style inconsistency types for models?
    \end{itemize}
    \item \textbf{RQ3:} 
     How do model-generated code and human-written code compare in three aspects: readability, conciseness, and robustness?
    \item \textbf{RQ4:} 
    Can prompting techniques improve the coding style of LLMs?
\end{itemize}

\subsection{RQ1: Coding Style Inconsistency Identification}
\label{sec:rq1}
To identify the inconsistencies in coding styles between LLMs and human programmers, we manually analyze the generated results of the five LLMs. It's worth noting that human-written code does not always equate to good coding style. \textbf{Therefore, we do not use human-written code as a standard to assess the quality of coding style in LLM-generated code. Instead, we focus solely on analyzing the coding style inconsistencies between human-written code and LLM-generated code.}
By comparing these generated results with the ground truths, we summarize the types of coding style inconsistencies. We conduct open coding~\cite{khandkar2009open} on the code generated by LLMs. Initially, we describe the data collection process, followed by a detailed explanation of the coding protocol. 

\subsubsection{Data Collection} 
Our data collection process includes three steps: model generation, automatic filtering, and manual filtering.

\textbf{Model generation.} 
For each of the 230 Python code generation tasks from CoderEval~\cite{yu2024codereval}, 
we prompt the five LLMs %(introduced in Section~\ref{sec:setting:models}) 
to perform code generation using the same prompting template. For each task, we instruct each model to generate 10 results, resulting in an initial total of 2,300 code samples for each model.

\textbf{Automatic filtering.}
To ensure the correctness of the collected code samples, we further filter out code samples that fail to pass any of the associated unit tests for the task, leading to 456, 189, 365, 497 and 570 results that pass all tests for five LLMs, respectively. We further merge identical code samples to reduce analysis effort, resulting in 1,557 unique code samples.
We only annotate 
% 1159 
these unique code samples to ensure that the annotation results for the same code sample generated by different models are consistent, thereby avoiding the situation where the same code sample generated by different LLMs is annotated with different results.

\begin{figure}[t]
\centering
\vspace{-3mm}
\begin{minipage}{0.45\columnwidth}
    \centering
    \includegraphics[width=\columnwidth]{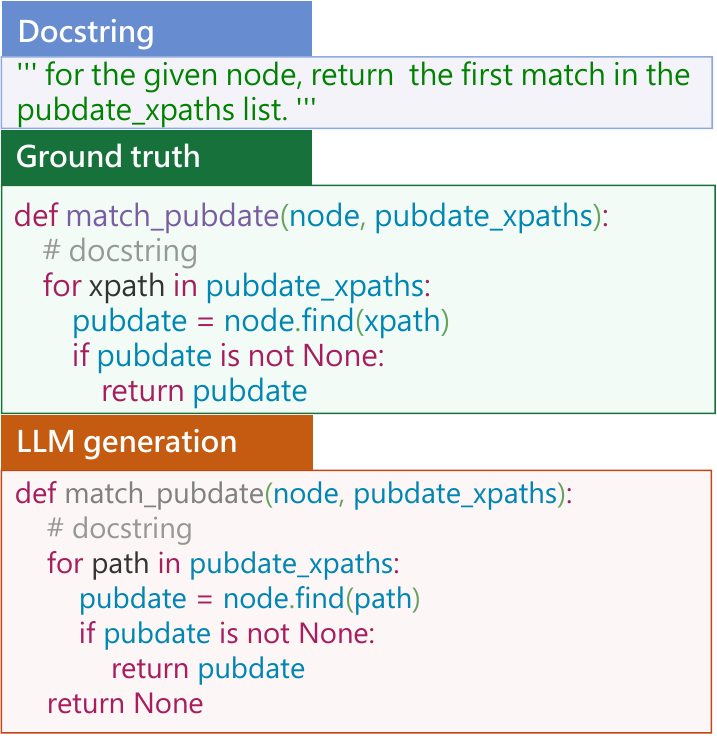}
    \figmargin
    \caption{An Example of Style-Consistent Implementation.}
    \label{fig:consistent_style_example}
\end{minipage}%
\hfill
\begin{minipage}{0.45\columnwidth}
    \centering
    \includegraphics[width=\columnwidth]{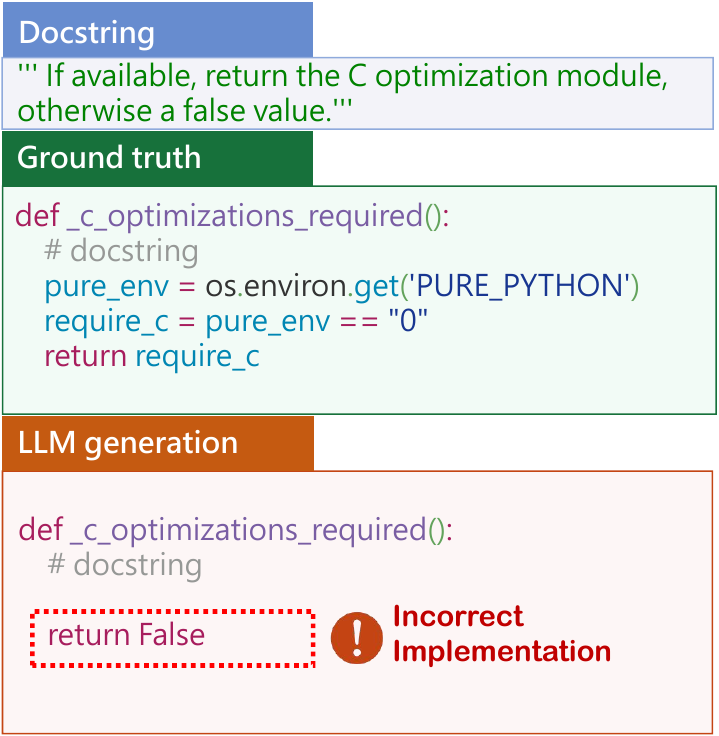}
    \figmargin
    \caption{An Example of Incorrect Implementation that Passed Unit Tests.}
    \label{fig:incorrect_implementation_example}
\end{minipage}
\end{figure}

\textbf{Manual filtering.}
To ensure the quality of collected code samples, we manually check and filter them based on three criteria: 
\blackcircled{1} \textit{Style consistency}. We filter out results that exhibit no inconsistency in coding style. 
%% 需要修改
If the naming conventions, commenting style, code structure, and other aspects of the two code samples are consistent, we conclude that there is no significant difference in code style between them.
For example, Figure~\ref{fig:consistent_style_example} shows an example of consistent coding style between the code sample generated by Code LLM and corresponding ground truth. As a result, 
% 56+17
73 code samples are filtered out in this way.
\blackcircled{2} \textit{Functional correctness.}  We filter out results that implement the task incorrectly despite passing the unit tests. Previous work has shown that existing benchmarks suffer from test sufficiency issues, meaning that even if a generated result passes all tests, there is still a chance it could be incorrect~\cite{EvalPlus}. For example, Figure~\ref{fig:incorrect_implementation_example} shows an example of wrong implementation although passing test cases. As a result, 
% 264
286 code samples are filtered out in this way.
\blackcircled{3} \textit{Implementation conciseness}. We filter out results that contain extra code that does not contribute to fulfilling the function's implementation requirements (e.g., two exactly the same loops). As a result, 19 code samples are filtered out in this way. Finally, we obtain 1,179 unique code samples for the study.

% \begin{mdframed}[backgroundcolor=yellow!25,hidealllines=true,innerleftmargin=0pt,innerrightmargin=0pt,innertopmargin=0pt,innerbottommargin=0pt, skipabove=0pt, skipbelow=0pt]
% The details of the annotation process need to be described in 4.1.2:

% (1) All three annotators have expertise in code style and experience in Python programming.

% (2) What is the rate of disagreement in annotation? 

% (3) How to determine that the two pieces of code do not have code style differences?

% (4) An example can be given to illustrate what happens when there is disagreement among the annotators (e.g., annotator A's annotation results in 1,3,5, and annotator B's annotation results in 1,3,4) and how to discuss to reach a consistent result (e.g., eventually, A and B are consistent after iteration, and the annotation results are 1,3,4)

% \end{mdframed}

\subsubsection{Data Annotation}
We adopt classifications of coding style inconsistencies in previous work~\cite{chen2023duetcs} as the initialization of our classification and conduct open coding~\cite{khandkar2009open} on the code samples generated by LLMs. 
Our objective is to refine and expand these classifications to capture detailed instances of coding style inconsistencies. \textbf{In our annotation process, three annotators were involved, each with over five years of Python programming experience and a solid understanding of the field of code style.}

% \textbf{Iterative coding.} We analyze the code samples one by one. For each code sample, we compare it with the ground truth line by line to identify the inconsistencies, without knowing which model produced the result. If a code sample and its corresponding ground truth show inconsistency that matches a current definition of inconsistency type, we code the generated result with the specific inconsistency type. If the inconsistency does not fit any existing definitions, we either modify an existing definition or create a new type. When the inconsistency types are updated, all code samples will be re-annotated to ensure consistency. 
% Note that a code sample can be classified under multiple inconsistency types. For example, if a code sample uses a different naming convention (Naming Formatting Inconsistency) and also structures loops differently (Loop Structure Inconsistency), it will be annotated with both inconsistency types.

\textbf{Iterative coding.} The three annotators independently analyze the code samples one by one. For each code sample, the annotators independently compare it with the ground truth line by line to identify inconsistencies, without knowing which model produced the result. If a code sample and its corresponding ground truth show inconsistency that matches a current definition of inconsistency type, the annotators code the generated result with the specific inconsistency type. If the inconsistency does not fit any existing definitions, the annotators either modify an existing definition or create a new type. When the inconsistency types are updated, all code samples will be re-annotated to ensure consistency. Note that a code sample can be classified under multiple inconsistency types. For example, if a code sample follows a different naming convention (Naming Formatting Inconsistency) and also structures loops differently (Loop Structure Inconsistency), it will be annotated with both inconsistency types.

This iterative coding process aims to capture the nuanced nature of coding style inconsistencies. During the coding process, the annotators also summarize guidelines for each inconsistency type annotation to ensure clarity and consistency in our annotations. These guidelines include specific examples and detailed descriptions to help identify and classify each type of inconsistency accurately. This ensures the annotation consistency and the reproducibility across different coders.

\textbf{Periodic review and update.} 
After analyzing every 50 code samples, the three annotators conduct a review of both the taxonomy and the coded samples. Based on insights from the review and discussions, we refine the definitions of inconsistency types, merging or removing types as necessary. Following any updates to the taxonomy, all code samples are re-annotated to maintain consistency and accuracy in the categorization of inconsistencies. This periodic review and update process continues until all code samples have been fully coded, ensuring thorough and reliable identification of coding style inconsistencies. Note that the taxonomy has remained stable during the last several reviews, indicating a mature and robust classification system. 
% Three of the authors perform the manual filtering and the coding together, resolving disagreements through discussions. 
We calculated that the average disagreement rate among the three annotators was 37.6\% after labeling every 50 samples. However, these disagreements were resolved through discussions among the annotators.

\subsubsection{Taxonomy}
Table~\ref{table:taxonomy:new1} presents the 24 inconsistency types identified during the open coding, along with their names and definitions. For each inconsistency type, the full annotation results and detailed annotation guidelines are included in our replication package~\cite{replication_package}. 

\textbf{Taxonomy Analysis.}
We have further categorized the 24 types of inconsistencies into five dimensions based on their main focus: 
\begin{itemize}%[leftmargin=10pt]
    \item \textbf{Formatting Inconsistency}. This dimension focuses on inconsistencies related to code formatting, such as indentation, spacing, and code/comment layout. 
    \item \textbf{Semantic Inconsistency}. This dimension focuses on inconsistencies related to the meaning or semantics of code, including variable naming, function naming, and the level of detail in comment style. 
    \item \textbf{Expression/Statement Inconsistency}. This dimension focuses on inconsistencies related to the style or usage of expressions and statements within the code, such as assignment styles, conditional expressions, and data structure construction.   
    \item \textbf{Control Follow Inconsistency}. This dimension focuses on inconsistencies related to control flow structures within the code, such as conditional statements, loop structures, and exception handling.   
    \item \textbf{Fault Tolerance Inconsistency}. This dimension focuses on inconsistencies related to error handling and fault tolerance mechanisms within the code, including input validation, runtime validation, and exception handling.  
\end{itemize}

It is worth noting that we have categorized 24 inconsistency types into 5 dimensions, with the scope of code style inconsistencies contained in each dimension solely determined by the inconsistency types assigned to that dimension. Since there is no obvious overlap between the inconsistency types, there is also no clear overlap between the 5 dimensions.

\begin{figure}[]
\centering
\vspace{-5mm}
\includegraphics[width=1.0\linewidth]{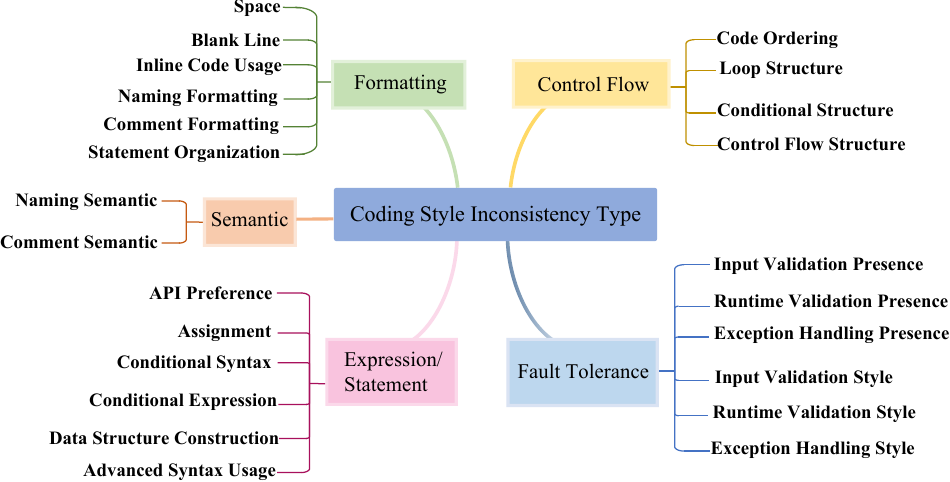}
\caption{Dimensions and Corresponding Coding Style Inconsistency Types.} 
\label{fig:taxonomy_relations}
\end{figure}

Figure~\ref{fig:taxonomy_relations} provides a visual representation of the relationships between the five dimensions and the 24 inconsistency types identified.
The inconsistency types are organized into a tree-like structure in the figure, with the dimensions and inconsistency types represented using different shapes, connected by lines. 
% Those inconsistency type sharing the same color indicate they belong to the same dimension. 
Furthermore, these inconsistencies vary in their scopes of influence, such as identifier, statement, and block, as also depicted in Figure~\ref{fig:taxonomy_relations}. Some inconsistencies may belong to only one or a few identifiers (e.g., Naming Formatting Inconsistency) or a single statement (e.g., Assignment Style Inconsistency), while  others may impact an entire block of code (e.g., Loop Structure Inconsistency) or span across multiple blocks (e.g., Code Order Inconsistency).

\begin{table}[H]
    \centering
    \footnotesize
    \scriptsize 
    \caption{Coding Style Inconsistency taxonomy.}
    \label{table:taxonomy:new1}
    \begin{tabularx}{\textwidth}{|c|c|X| }
        \hline
        \textbf{ID} & \textbf{Inconsistency Type} &  \textbf{Definition} \\ \hline
         1 & \NamingFormat  Inconsistency & Inconsistencies in the formatting of identifiers (e.g., variable names, function names, or parameter names), such as using camelCase (e.g., authorName) versus snake\_case (e.g., author\_name). \\ \hline

         2 & \Space Inconsistency  &  Inconsistencies in the use of space(e.g., whitespace and indentation)  around various syntactical elements, e.g., operators, colons, comments, and brackets. \\ \hline

         3 & Blank Line Inconsistency   &  Inconsistencies in the use of blank lines. For example, one style includes blank lines to separate code blocks, while the other omits them. \\ \hline

         4 & Inline Code Usage Inconsistency  & Inconsistencies in the usage of inline code constructs. It encompasses cases where one approach employs inline expressions or functions while the other does not. \\ \hline

         5 & Comment Format Inconsistency  & Inconsistencies in the formatting of comments within code. It includes variations in interline comments, inline comments, commented-out code, and trailing comments. \\ \hline

         6 & Statement Organization Inconsistency & Inconsistency in the organization style of statements, exemplified by completing expressions or statements in a single line in contrast to breaking them into multiple shorter lines. \\ \hline

         7 & \NamingSemantics Inconsistency  & Inconsistencies in the semantic meaning of identifiers, such as using generic single-letter identifiers (e.g., i, l, d) versus meaningful, descriptive words (e.g., index, length, day).    \\ \hline
         
         8 & Comment Semantics Inconsistency  & Inconsistencies in the semantic aspects of comments within code, such as variations in the level of detail or semantic differences, or with TODO comment, useless comments. \\ \hline

         9 & Assignment Inconsistency & Inconsistencies in the style of variable assignment, e.g., tuple unpacking assignment, chained assignment, separate assignment. Examples include using augmented assignment versus standard assignment (e.g., `x += 1` vs. `x = x + 1`).  \\ \hline

         10 & Conditional Syntax Inconsistency  & Inconsistencies in the syntax used for conditional statements within code. It covers scenarios where one method involves conditional statements while the other employs conditional expressions or return statements with equivalent functionality.\\ \hline

        11 &  Conditional Expression Inconsistency & Inconsistencies in the way conditional expressions are written, despite having similar functionalities. For example, one style might use if len(a) > 1 while another uses if len(a) >= 2. \\ \hline

        12 & Data Structure Construction Inconsistency  & Inconsistencies in the methods used to construct data structures such as lists, dictionaries, sets, tuples, strings, and iterators. For example, using different syntaxes or functions to create these data structures. \\ \hline
         
        13 & API Preference Inconsistency & Inconsistencies in how APIs are used to achieve similar functionality. It includes variations such as calling different functions or methods defined in the repository, using built-in functions, or re-implementing the functionality without calling existing functions. \\ \hline

        14 & Advanced Syntax Usage Inconsistency  & Inconsistencies in the use of advanced syntax features, such as lambda expressions \\ \hline

        15 & Code Ordering Inconsistency  & Inconsistencies in the order of semantically similar code blocks, such as import statements, assignments, loops, and other logical sections of code.\\ \hline

        16 &  Loop Structure Inconsistency  & Inconsistencies in loop structures within code. It covers scenarios where one approach employs a for loop while the other uses a while loop, or where one loop contains only a basic loop structure while the other includes additional control flow statements such as if-break, for-else structure, and while-else structure \\ \hline
        
        17 & Conditional Structure Inconsistency  & Inconsistencies in the structure and design of conditional statements within methods. It includes variances such as the use of multiple conditional statements versus a single statement with equivalent meaning, differences in the structures of multiple conditional statements while preserving the same semantics, and disparities in the inclusion of return statements alongside conditional statements. \\ \hline
        
        18 & Control Flow Structure Inconsistency  & Inconsistencies in the use of control flow structures, such as using if-else versus try-except statements during the code execution (e.g., input and runtime validation).   \\ \hline

        19 & Input Validation Presence Inconsistency  & Inconsistencies in whether input checking with conditionals is performed. \\ \hline

        20 &  Runtime Validation Presence Inconsistency  & Inconsistencies in whether runtime validation with conditionals is performed, ensuring data integrity during code execution.. \\ \hline
        
        21 & Exception Handling Presence Inconsistency  & Inconsistencies in whether exceptions are handled, e.g., using try-except blocks, to manage errors that occur during execution.  \\ \hline

        22 & Input Validation Style Inconsistency  & Inconsistencies in the style of input validation with conditionals (ensuring input data is checked before processing), such as whether exceptions are thrown, the types of exceptions used, and the use of logging. \\ \hline

        23 & Runtime Validation Style Inconsistency & Inconsistencies in the style of runtime validation with conditionals during code execution, such as whether exceptions are thrown, the types of exceptions used, and the use of logging.\\ \hline
        
        24 & Exception Handling Style Inconsistency   & Inconsistencies in the style of exceptions that occur during execution are handled, such as whether exceptions are thrown, the types of exceptions used, and the use of logging, the use of try-else block. \\ \hline
   \end{tabularx}
\end{table}

% \begin{mdframed}[backgroundcolor=yellow!25,hidealllines=true,innerleftmargin=0pt,innerrightmargin=0pt,innertopmargin=0pt,innerbottommargin=0pt, skipabove=0pt, skipbelow=0pt]
% (1) Add content: We picked a subset and wrote ground truths by hand, and found that the category table did not change significantly. (There is only one ground truth for each task in CoderEval, each time the model generated code is compared with this ground truth, does it mean that each comparison considers only one human programmer's code style?)
% \end{mdframed}

% \begin{figure}[t]
% \centering
% %\vspace{-5mm}
% \includegraphics[width=1.0\linewidth]{images/dimensions_and_types.pdf}
% \figmargin
% \caption{Dimensions and Corresponding Coding Style Inconsistency Types.} 
% \label{fig:taxonomy_relations}
% \figmargin
% \end{figure}

Note that certain inconsistencies could affect both statement and block structures, contingent upon the complexity of the code involved. For instance, in the context of API preference inconsistency, the implementation of the same functionality may vary. It could involve calling different single APIs within a statement, or it might require the coordination of several APIs with specific usage patterns across multiple code blocks.

\textbf{Taxonomy Comparison.} Compared with the coding style taxonomy of Chen et al.~\cite{chen2023duetcs}, they categorize coding styles into text style and structure style, with four subtypes formatting, naming, ordering of code blocks, and control structures. Our taxonomy covers all these types and introduces three additional dimensions: semantic, expression/statement, and fault tolerance. We expand upon their framework by introducing 24 fine-grained types compared to 4 types.
For instance, we refine their subtype Control Structures into three specific inconsistency types related to: Conditional Structure Inconsistency, Loop Structure Inconsistency, and Control Flow Structure Inconsistency, offering a more detailed classification. Our taxonomy is backed by comprehensive guidelines derived from actual open coding, providing detailed and actionable classifications.

\textbf{Taxonomy Generalizability.} One concern is that, since each task in CoderEval has only one ground truth, this may reflect the coding style of a single programmer, potentially limiting the generalizability of our taxonomy. To this end, we randomly selecting 42 Python tasks from the 230 available in CoderEval and two annotators were assigned to independently rewrite the ground truths for these 42 tasks, resulting in a total of 84 ground truths. Another annotator was responsible for verifying the correctness of these ground truths. We then identified the code samples corresponding to these 42 tasks from a set of 1,179 unique code samples and annotated the coding style differences between the rewritten ground truths and the code samples following the outlined annotation process. Both the rewritten ground truths and the annotation results are available in our replication package~\cite{replication_package}. After the annotation process, the overall taxonomy remained unchanged.

In summary, our taxonomy not only complements but also substantially enhances previous research, filling critical gaps and offering a more robust framework for analyzing the inconsistencies in coding style. Note that while our taxonomy is based on summarizing inconsistencies observed in Python code generated by LLMs, it is not limited to Python alone. The concepts and categories can be generalized to other programming languages as needed.

\begin{center}
    % \begin{myboxb}[]{RQ3 Summary} %ab
    \begin{myboxc} \textbf{RQ1 Summary: } %cd
    % \begin{myboxd} \textbf{RQ4 Summary: } %cd
   We have identified 24 types of coding style inconsistencies and categorized them into five dimensions: Formatting, Semantic, Expression/Statement, Control Flow, and Fault Tolerance. Our taxonomy expands upon previous work by introducing new dimensions and providing more detailed classifications with guidelines. 
    \end{myboxc} %cd
    % \end{myboxd} %cd
    % \end{myboxb} %ab
\end{center}

\subsection{RQ2: Coding Style Inconsistency Analysis}
\label{sec:rq2}
We design RQ2 to evaluate the differences between human-written code and LLM-generated code. Specifically, we investigate the coding style differences in three perspectives: (1) Percentages of inconsistent coding styles; (2) Inconsistency numbers present in a single code sample; and (3) Distribution of coding style inconsistency types.

% \begin{figure}[t]
% \centering
% \includegraphics[width=1.0\columnwidth]{images/no_style_inconsistent_vs_style_inconsistent.pdf}
% \figmargin
% \caption{Percentages of Inconsistent Coding Styles. DSC-1.3B and DSC-6.7B are short for \deepsmall and \deepbig, respectively.}
% \label{fig:distribution:inconsistency}
% \figmargin
% \end{figure}

\begin{figure}[t]
\centering
\begin{minipage}{0.48\columnwidth}
    \centering
    \includegraphics[width=\columnwidth]{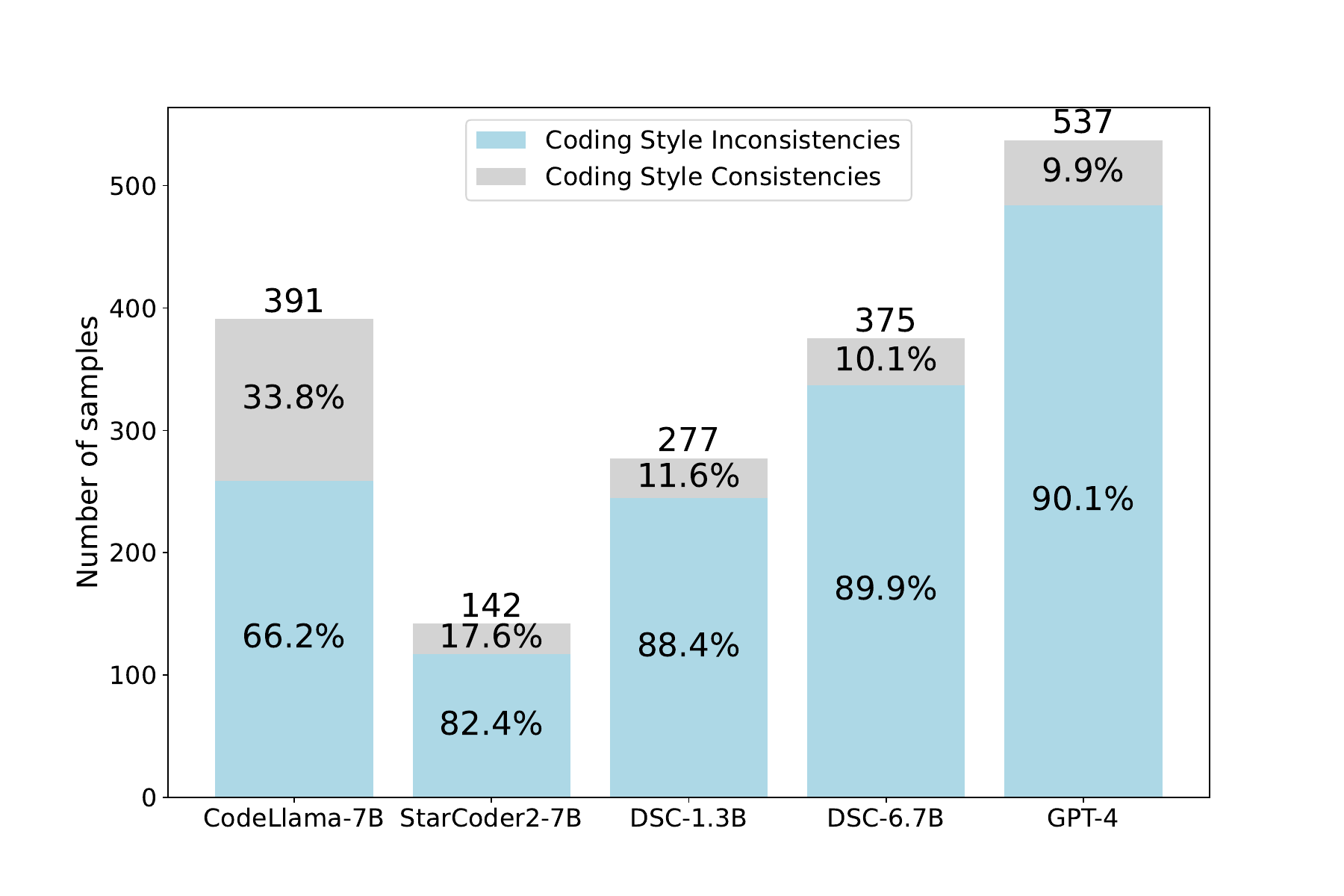}
    \figmargin
    \caption{Percentages of Inconsistent Coding Styles. }
    %DSC-1.3B and DSC-6.7B are short for \deepsmall and \deepbig, respectively.
    \label{fig:distribution:inconsistency}
\end{minipage}%
\hfill
\begin{minipage}{0.48\columnwidth}
    \centering
    \includegraphics[width=\columnwidth]{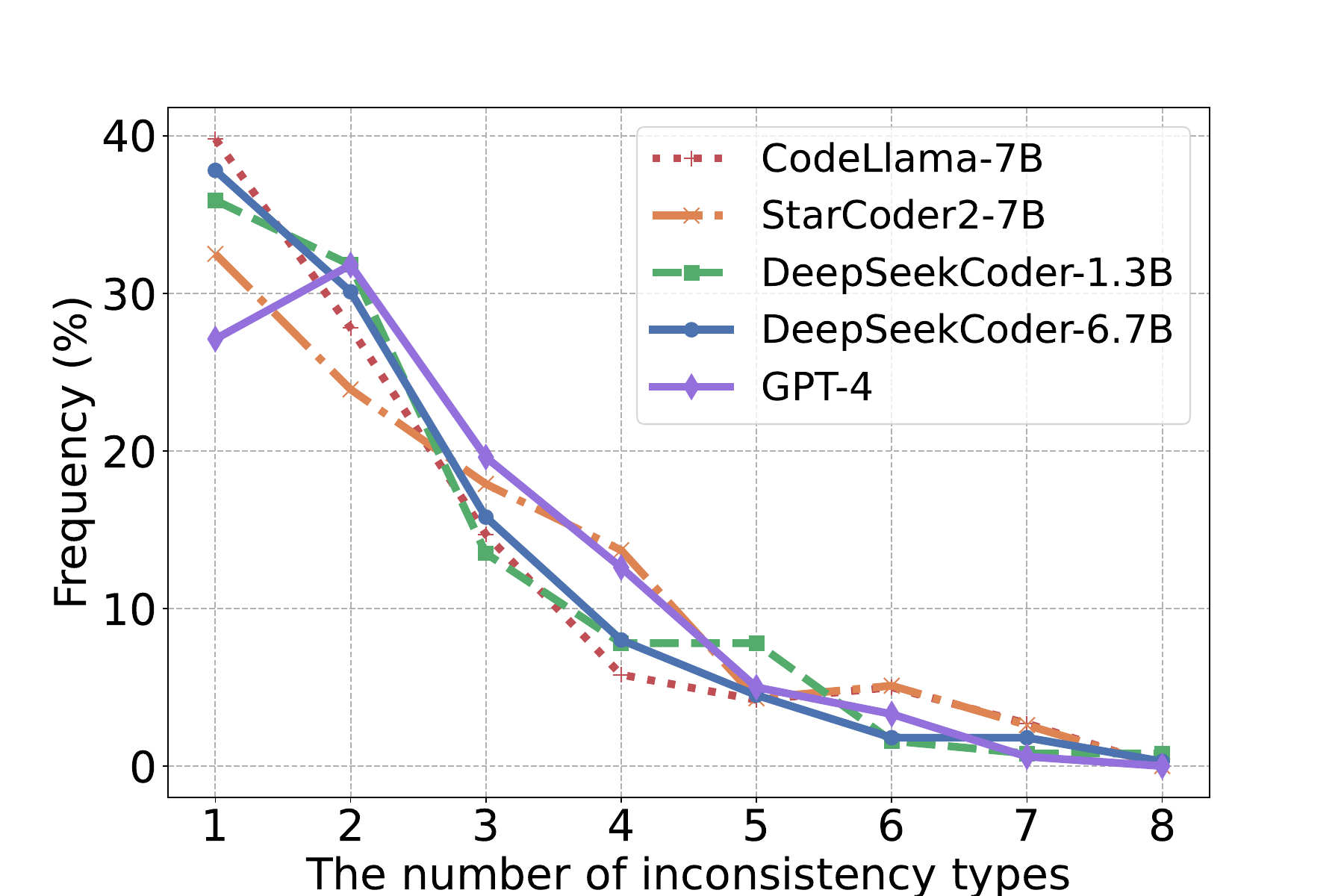}
    \figmargin
    \caption{Inconsistency Numbers in a Single Code Sample.}
    \label{fig:numbersoftypes}
\end{minipage}
\end{figure}

\subsubsection{Percentages of Inconsistent Coding Styles}

% \begin{mdframed}[backgroundcolor=yellow!25,hidealllines=true,innerleftmargin=0pt,innerrightmargin=0pt,innertopmargin=0pt,innerbottommargin=0pt, skipabove=0pt, skipbelow=0pt]
% (1) It is necessary to elaborate on how 391, 142, 277, and 375 came to be
% \end{mdframed}

% We first analysis the annotation of the code samples generated by Code LLMs from the presence or absence of coding style inconsistencies. 
% Figure~\ref{fig:distribution:inconsistency} shows the percentages of inconsistent coding styles for each Code LLM highlighting the presence or absence of coding style inconsistencies. 
% The initial number of functionally correct code samples (before deduplication) produced by the five Code LLMs (\codellama{}, \starcoder{}, \deepsmall{}, \deepbig{} and \gptfour{}) are 391, 142, 277, 375 and 537, respectively.

We counted the number of code samples generated by the five models (\codellama{}, \starcoder{}, \deepsmall{}, \deepbig{}, and \gptfour{}) that passed unit tests, were manually verified for functional correctness, and did not contain extra code. The results were 391, 142, 277, 375, and 537, respectively. Furthermore, we calculated the proportion of these code samples that either exhibit or do not exhibit coding style inconsistencies compared to human-written code.  The statistical results are shown in Figure~\ref{fig:distribution:inconsistency}. From Figure~\ref{fig:distribution:inconsistency}, we can find that all LLMs exhibit coding style inconsistency with human-written code and the degree varies: 66.2\%, 82.4\%, 88.5\%, 89.9\% and 90.1\% for the five models, respectively.

\subsubsection{Inconsistency Numbers Present in a Single Code Sample}

% \begin{figure}[t]
% \centering
% \includegraphics[width=0.8\columnwidth]{images/figure3.pdf}
% \figmargin
% \caption{Inconsistency Numbers in a Single Code Sample.} 
% \label{fig:numbersoftypes}
% \figmargin
% \end{figure}

% \textbf{Analysis on number of Inconsistency Types.} 
For each model, we counted the number of inconsistency types present in each code sample (considering only the code samples that exhibit coding style differences with the corresponding ground truths). Then, We counted the frequency of different numbers of inconsistent types in one sample for each model. A line chart was plotted based on the frequency of inconsistency types present in the code samples. From Figure~\ref{fig:numbersoftypes}, it can be seen that the number of inconsistent types for one code sample ranges between 1 and 8. For each model, the trend of the frequency line chart is roughly the same, with all lines generally showing a decreasing trend. 
% Among them, the code samples of the models all have the highest frequency of having 1 inconsistency type, at 34\%, 28\%, 38\%, and 37\% respectively. 
The code samples of \codellama{}, \starcoder{}, \deepsmall{}, and \deepbig{} show the highest frequency of having one inconsistency type, at 39.8\%, 32.5\%, 35.9\%, and 37.8\%, respectively. The code samples of \gptfour{} exhibit the highest frequency of two inconsistency types, at 31.8\%. The lowest frequency is that code samples with 8 inconsistency types, at 0.0\%, 0.0\%, 0.8\%, 0.3\% and 0.0\%, respectively.

\subsubsection{Distribution of Coding Style Inconsistency Types}

\vspace{5pt}
\begin{figure*}[t]
\centering
\begin{minipage}[t]{0.99\linewidth}
\centering
\includegraphics[width=0.63\linewidth]{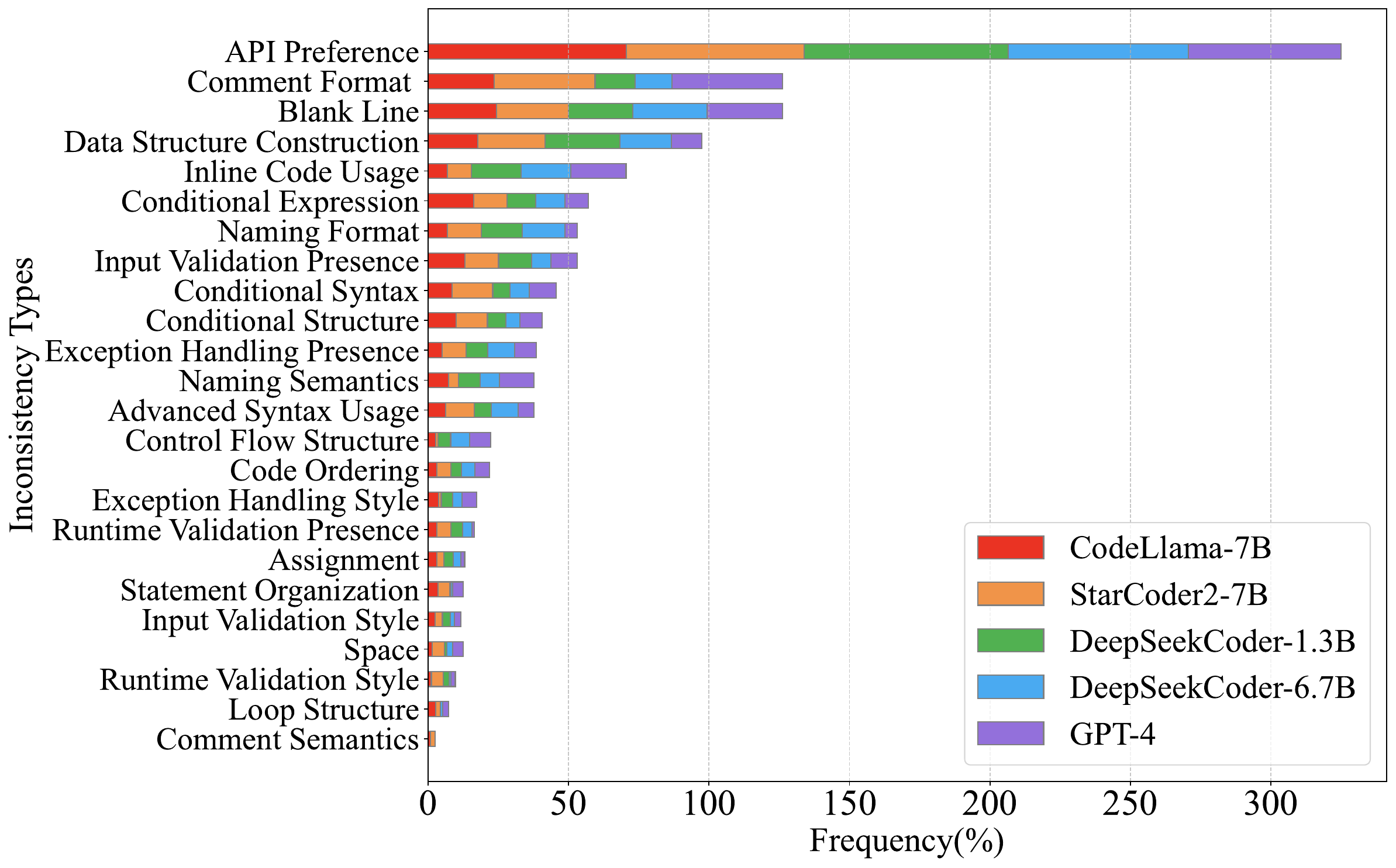}
\figmargin
\caption{Overall Distribution of Coding Style Inconsistency Types.}
\figmargin
\label{fig:distribution:overall}
\end{minipage}%
\vspace{20pt}
\vfill  

\begin{minipage}[t]{0.54\linewidth}
\centering
\includegraphics[width=\linewidth]{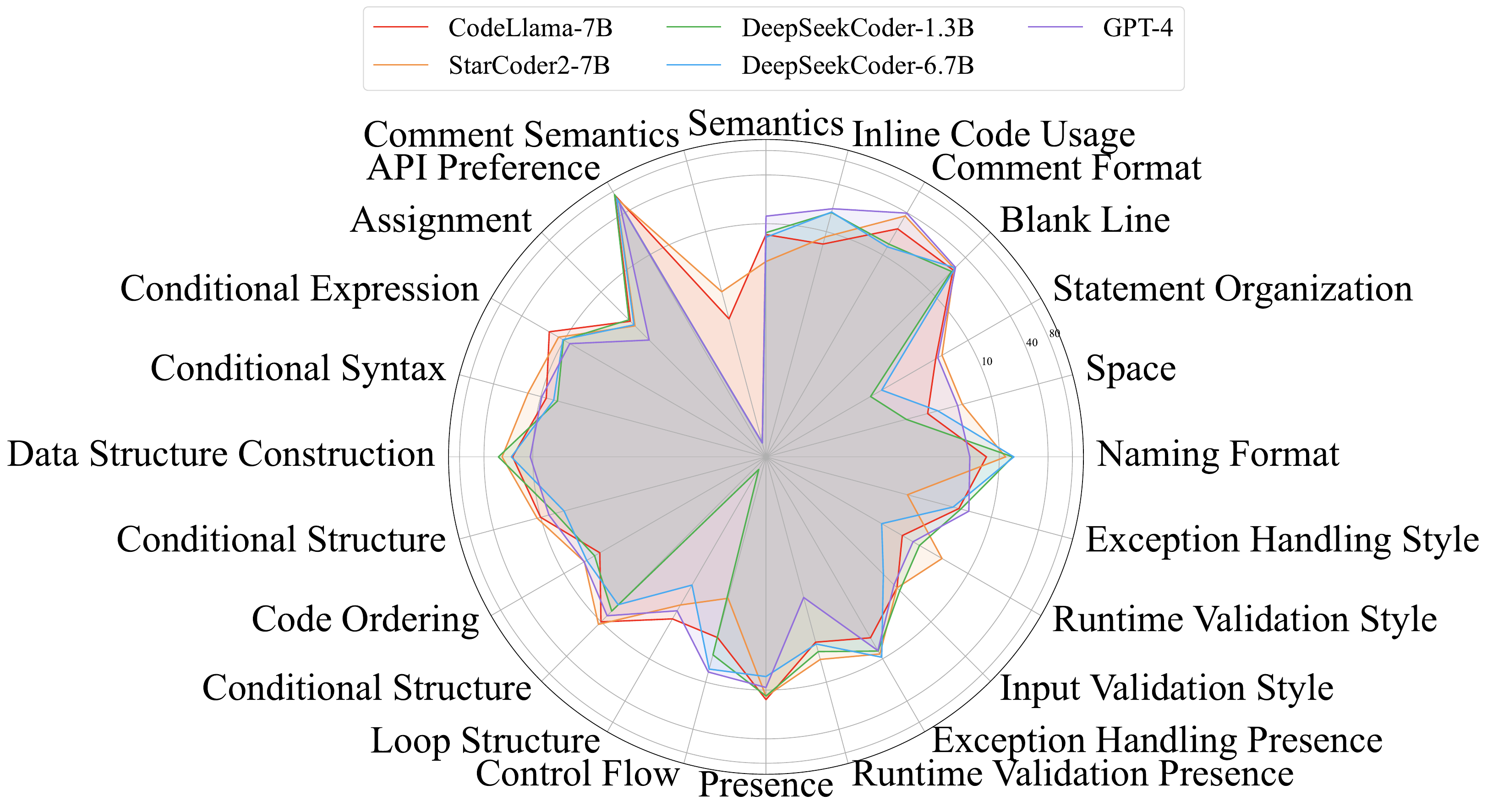}
\figmargin
\caption{Breakdown Distribution of Coding Style Inconsistency (by Model). The ``Semantics'' at the top refers to ``Naming Semantics'', and the ``Presence''at the bottom refers to ``Input Validation Presence''.}
\figmargin
\label{fig:distribution:byinconsistencytype}
\end{minipage}%
\hfill
\begin{minipage}[t]{0.45\linewidth}
\centering
\includegraphics[width=\linewidth]{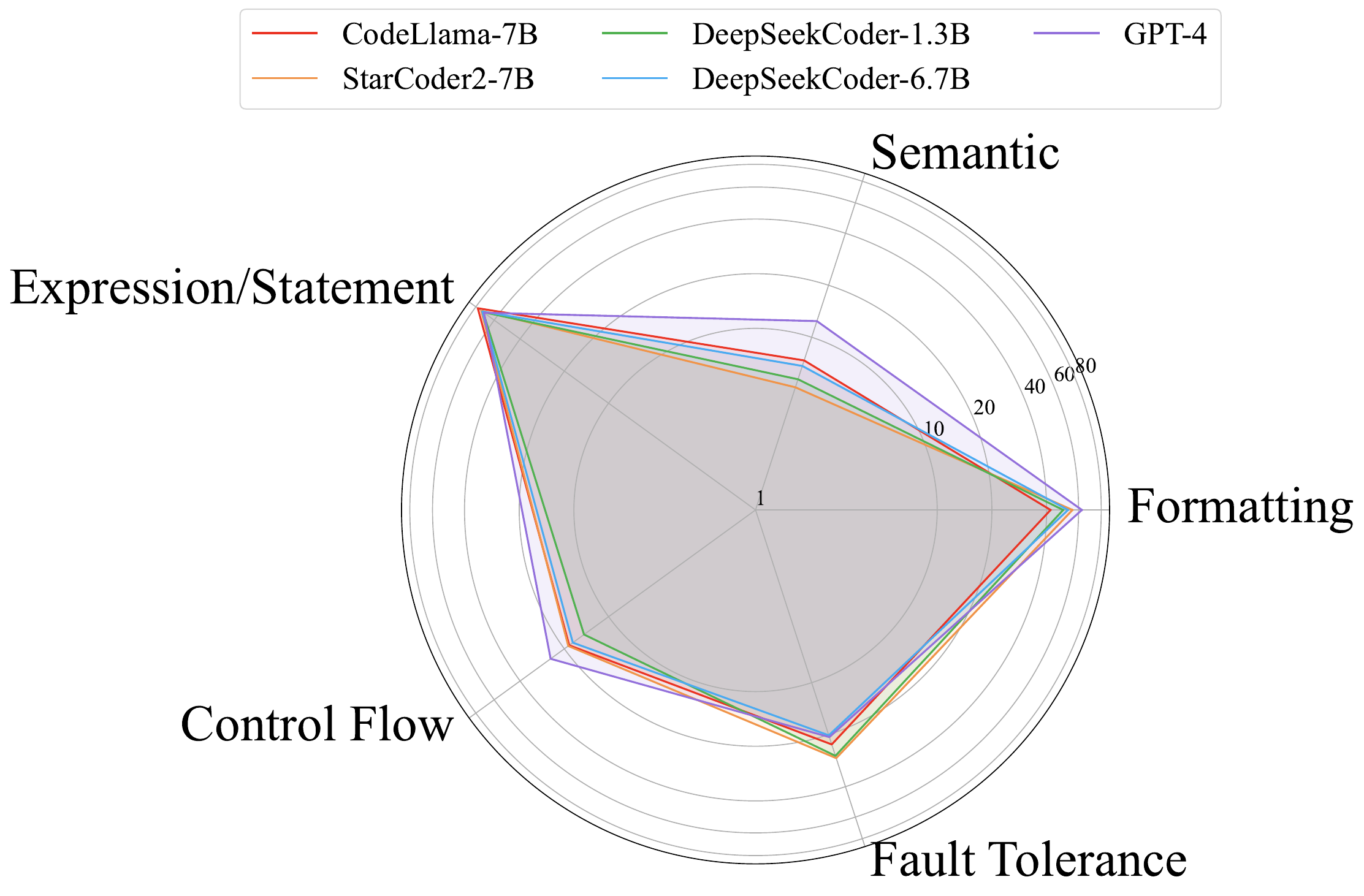}
\figmargin
\caption{Breakdown Distribution of Coding Style Inconsistency (by Dimension).}
\figmargin
\label{fig:distribution:bydimension}
\end{minipage}%
\end{figure*}

% Figure~\ref{fig:distribution:overall} illustrates the overall inconsistency distribution in different models. 
% We can observe that the top-4 inconsistency types are API Usage (270.7\%), Blank Line (99.2\%), Comment Formatting  (86.8\%), and Data Structure Construction (86.6\%),  significantly higher than other inconsistency types. 
% Among these top four inconsistency types, API Usage Inconsistency stands out with a significantly higher frequency, even surpassing the combined frequencies of the second and third-ranked types. 
% In contrast, the bottom inconsistency types are: Comment Semantics Inconsistency, Loop Structure Inconsistency, Runtime Validation Inconsistency, Space Inconsistency, Statement Organization Inconsistency, and Input Validation Inconsistency. The low frequencies in these types indicate that Code LLMs and human-written code are relatively consistent in these aspects. 

% Figure~\ref{fig:distribution:overall} illustrates the overall inconsistency distribution in these models. For example

Figure~\ref{fig:distribution:overall} is a stacked plot showing the frequencies of a certain inconsistency type in code samples generated by five models. For example, the frequencies of API Preference inconsistency in the code samples generated by \codellama{}, \starcoder{}, \deepsmall{}, \deepbig{}, and \gptfour{} are 70.7\%, 63.3\%, 72.7\%, 64.1\%, and 54.4\%, respectively, summing up to 325.0\%.
We can observe that the top-4 inconsistency types are API Preference Inconsistency (325.0\%), Comment Formatting Inconsistency (126.3\%), Blank Line Inconsistency (126.1\%) and Data Structure Construction Inconsistency (97.4\%). Among these top four inconsistency types, API Preference Inconsistency stands out with a significantly higher frequency, even surpassing the combined frequencies of the second and third-ranked types. In contrast, the bottom inconsistency types are: Runtime Validation Style (9.7\%), Loop Structure (7.4\%), Comment Semantic (2.5\%). The low frequencies of these types indicate that LLMs and human-written code are relatively consistent in these aspects. 

In order to understand the inconsistencies deeper, we conducted a detailed analysis of the top-4 inconsistency types. In our observed code samples and the ground truths, we found that the code samples and the ground truths might call functions from different sources to achieve similar functionality. Different sources refer to functions that may be defined within the original repository, built-in Python functions, etc. For example, we found that in 6.6\% of cases, the ground truth calls functions defined in the original repository while similar functionality is achieved using Python built-in functions, etc., in the code samples generated by models. This may be because the model lacks contextual information about the functions defined in the original repository when generating code. As a result, the large model uses built-in functions or third-party library functions, etc., to achieve similar functionality. For instance, in one task, the ground truth uses a function defined in the original repository, \textit{``match\_file\_by\_prefix(prefix, file)''}, to check if the prefix of the file name is ``prefix'', while the code sample generated by models uses the built-in method in Python \textit{``startswith''} to achieve similar functionality.

Comment Format Inconsistency and Blank Lines Inconsistency are the second and third most frequent inconsistency types. In our analysis, we found that model-generated code samples tend to avoid using blank lines to separate code blocks compared to human-written code.
% the explanation of comment formatting inconsistency
The five models generally have a high frequency of Comment Formatting inconsistency, but there are differences among them (The highest frequency is observed for \gptfour{} at 39.5\%. \starcoder{} follows with 35.9\%, and \codellama{} ranks third at 23.6\%. \deepsmall{} and \deepbig{} have the lowest frequencies, recorded at 14.3\% and 13.1\%, respectively.). 
% The reason for the high frequency of Comment Formatting Inconsistency across the five models is that, in our observed code samples and corresponding ground truths, the code generated by models shows a preference against generating semantically meaningful block comments compared to the code written by human programmers. 
% Among these models, \gptfour{} shows the highest frequency, partly because it tends to include more meaningful block comments and inline comments compared to human-written code. The other four models, in comparison to human-written code, are not inclined to write block comments and inline comments. The code samples generated by \codellama{}, \starcoder{}, and \deepsmall{} even do not include any examples containing inline comments. One reason for the relatively high frequency of \starcoder{} and \codellama{} is that the comment formatting in the code samples generated by \starcoder{} and \codellama{} is less standard compared to that in the code samples generated by the other models.
Among these models, \gptfour{} has the highest frequency, partly because it includes more meaningful block and inline comments compared to human-written code. The other four models are less inclined to add such comments. Notably, code samples from \codellama{}, \starcoder{}, and \deepsmall{} even do not include any examples containing inline comments. The relatively high frequency of \starcoder{} and \codellama{} can be attributed to their less standardized comment formatting compared to the other models. 
For example, the code samples generated by \codellama{} and \starcoder{} may contain commented-out code or TODO comments, while the code samples generated by \deepsmall{}, \deepbig{} and \gptfour{} do not. We consider that having commented-out code in code is not good coding practice because these comments are unnecessary information and do not help in understanding the functionality of the code. We believe that including TODO comments in code generated by large models is not good coding practice. This is because we expect large models to produce complete code based on requirements, rather than including comments indicating unfinished tasks or future improvements. Data structure construction inconsistency is a frequently occurring type of inconsistency. The code samples and the corresponding ground truths may show differences in constructing data structures (e.g.,  list, set). In our observed samples, human programmers tend to prefer using list comprehensions to construct lists, whereas the code samples generated by LLMs tends to favor conventional methods for constructing lists.

Figure~\ref{fig:distribution:byinconsistencytype} shows a radar chart of the frequency of inconsistency types for five different models, allowing us to compare the overall frequency distribution of inconsistency types across these models. 
As shown in Figure~\ref{fig:distribution:byinconsistencytype}, the distribution of inconsistency types for \deepsmall{} and \deepbig{} is relatively similar compared to the other models in terms of inconsistency types such as comment format, conditional expression, and naming format, etc. For example, in the Inline Code Usage inconsistency type, the frequency for \deepsmall{} and \deepbig{} is very close, lower than the frequency for \gptfour{}, but higher than that for \codellama{} and \starcoder{}. The frequencies of \deepsmall{}, \deepbig{}, and \gptfour{} are higher because they tend to include more intermediate variables in the code compared to the ground truths. Therefore, we can conclude that the base model significantly influences the coding style. The training data and method have a more noticeable impact on the coding style of the model compared to the parameters.

Figure~\ref{fig:distribution:bydimension} presents a radar chart that summarizes coding style inconsistencies by grouping them into five broader dimensions, i.e., formatting, semantic, expression/statement, control flow, and fault tolerance. 
To calculate the frequency for each dimension, we sum the instances of inconsistency types belonging to that dimension and divide it by the total number of valid code samples. 
From Figure~\ref{fig:distribution:bydimension}, we have the following observations:
\begin{itemize}%[leftmargin=10pt]
\item It is evident that the coding styles of different LLMs are similar in  dimension granularity. This is indicated by the almost overlapping shapes on the radar chart, highlighting that these models share a similar distribution of inconsistency types by dimension. 
\item The dimensions, ranked by average frequency of inconsistencies, are as follows: statement/expression (73.1\%), formatting (52.4\%), fault tolerance (24.3\%), control flow (18.0\%), and semantic(7.5\%). The high ranking of statement/expression inconsistency is primarily due to the significantly high frequency of API Preference Inconsistency within this dimension.  
\item We then calculate the difference between the highest and lowest values of frequency of inconsistencies for each dimension. We sort the five dimensions from high to low according to the difference, and the result is: formatting (20.5\%), semantic (7.3\%), statement/expression (7.2\%), fault tolerance (7.2\%), and control flow (6.0\%). This is because, although the training data of the models is generally similar, there are still some differences.
\end{itemize}

\begin{center}
    % \begin{myboxb}[]{RQ3 Summary} %ab
    \begin{myboxc} \textbf{RQ2 Summary: } %cd
    There are obvious coding style inconsistencies between human and all the studied LLMs. The top inconsistency type is API preference and top inconsistency dimensions are statements/expressions and formatting dimensions. While LLMs generally have similar coding styles, there are also noticeable differences in the formatting dimension.
    \end{myboxc} %cd
    % \end{myboxd} %cd
    % \end{myboxb} %ab
\end{center}

\subsection{RQ3: Coding Style Comparison}
\label{sec:rq3}

\begin{figure*}[t]
\centering
\begin{minipage}[t]{0.48\textwidth}
    \centering
    \includegraphics[width=\textwidth]{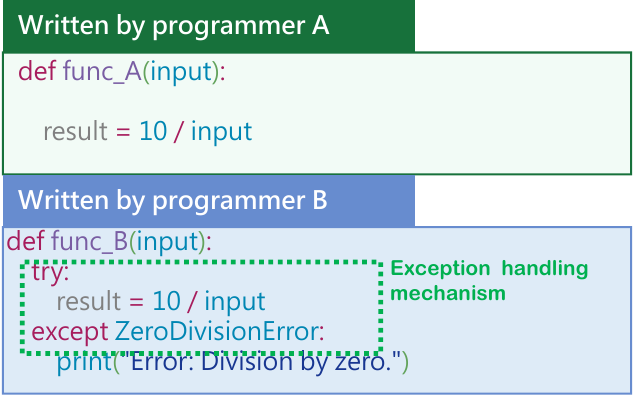}
    \caption{An example of how robustness relates to code style.}
    \label{fig:robustness_example}
\end{minipage}
\hfill
\begin{minipage}[t]{0.48\textwidth}
    \centering
    \includegraphics[width=\textwidth]{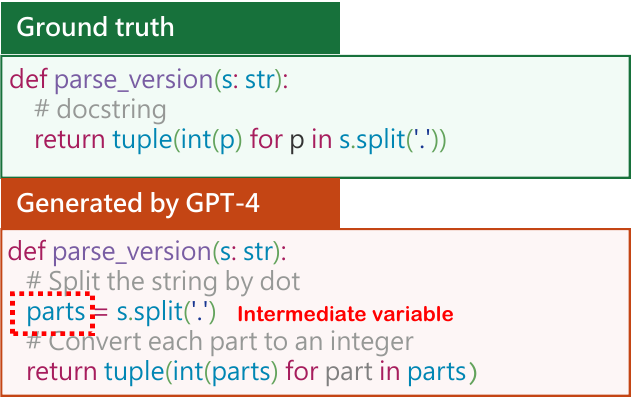}
    \caption{An example of generated code being less concise than Human-Written code.}
    \label{fig:conciseness}
\end{minipage}
\end{figure*}

% \begin{mdframed}[backgroundcolor=yellow!25,hidealllines=true,innerleftmargin=0pt,innerrightmargin=0pt,innertopmargin=0pt,innerbottommargin=0pt, skipabove=0pt, skipbelow=0pt]
% (1) To add the scoring guidelines. 
% (2) Explain why "robustness" is related to code style. “Furthermore, it is difficult to understand why the "robustness" metric is related to coding style. A correct implementation should correctly handle extreme cases and potential errors, and be consistent in its behavior on any input. If two functions behave differently for a given input, they are functionally different, not just stylistically inconsistent”.
% (3) Supplement the inner-outer agreement for RQ3 and RQ4, and explain that the results for RQ4 are statistically significant (calculate the p-value).
% \end{mdframed}

% In addition to the analysis of coding style inconsistency between Code LLMs and human programmers, we further investigate which coding style is better. To this end, we annotate the code generated by Code LLMs by comparing it with the ground truth from three aspects: readability, conciseness, and robustness. 

In addition to analyzing coding style inconsistency between LLMs and human programmers, we further compare the code samples generated by LLMs and the ground truths in terms of three aspects: readability, conciseness, and robustness. 
\begin{itemize}
    \item Readability: The readability and understandability of code. 
    \item Conciseness: The simplicity of the code and the degree to which it is free of unnecessary elements. 
    \item Robustness: 
    The ability of the code to handle corner cases and potential errors. To illustrate the relationship between robustness and code style, consider two programmers, A and B. Programmer B prefers to implement exception handling mechanisms in the code, while Programmer A tends to overlook this practice. In the Figure~\ref{fig:robustness_example}, \texttt{func\_A} works correctly when the input is valid, but it will crash if the input is zero. In contrast, \texttt{func\_B} includes an exception handling mechanism, ensuring the program does not crash and providing a meaningful error message when division by zero occurs. This demonstrates how  Programmer B's code is more robust due to different coding style.
\end{itemize}
\setlist[itemize]{noitemsep, topsep=0pt}
We compare the code samples generated by LLMs from RQ1 with their corresponding ground truths and evaluate each of the three aspects. For each analysis, we determine whether the generated code is better than the ground truth ("model better"), whether they are comparable ("tie"), or whether the ground truth is better ("human better"). The scoring for the three aspects follows these principles:
% model better (generated code is better than ground truth), tie (generated code is comparable to ground truth), and human better (the ground truth is better than the generated code).
(i) Readability: If code sample A conveys the intent and logical structure more clearly than code sample B, it is superior in readability.
(ii) Conciseness: If code sample A achieves the same functionality as code sample B while eliminating unnecessary redundancies, it is considered superior in terms of conciseness. This involves reducing the number of intermediate variables, optimizing loops, etc.
(iii) Robustness: If code sample A effectively implements error handling and fault tolerance mechanisms, whereas code sample B lacks such mechanisms, the former is deemed superior in robustness. This includes the presence of input validation, exception handling, etc. 

\begin{figure}[t]
\centering
\hspace{-10pt}
\subfigure[Readability]{
    \label{fig:rq3-readability}
    \includegraphics[width=0.33\linewidth]{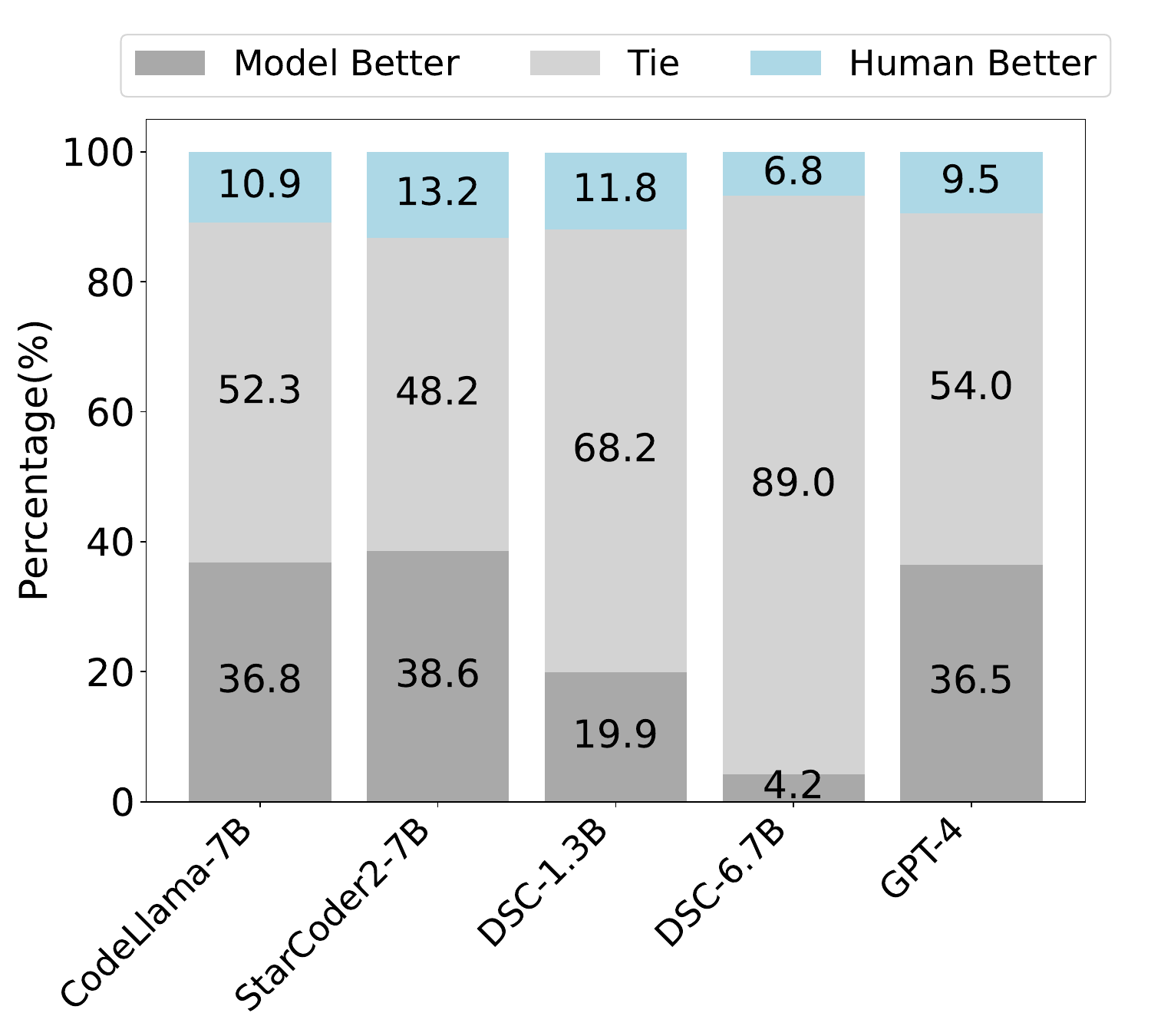}
}\hspace{-5pt}
\subfigure[Conciseness]{
    \label{fig:rq3-conciseness}
    \includegraphics[width=0.33\linewidth]{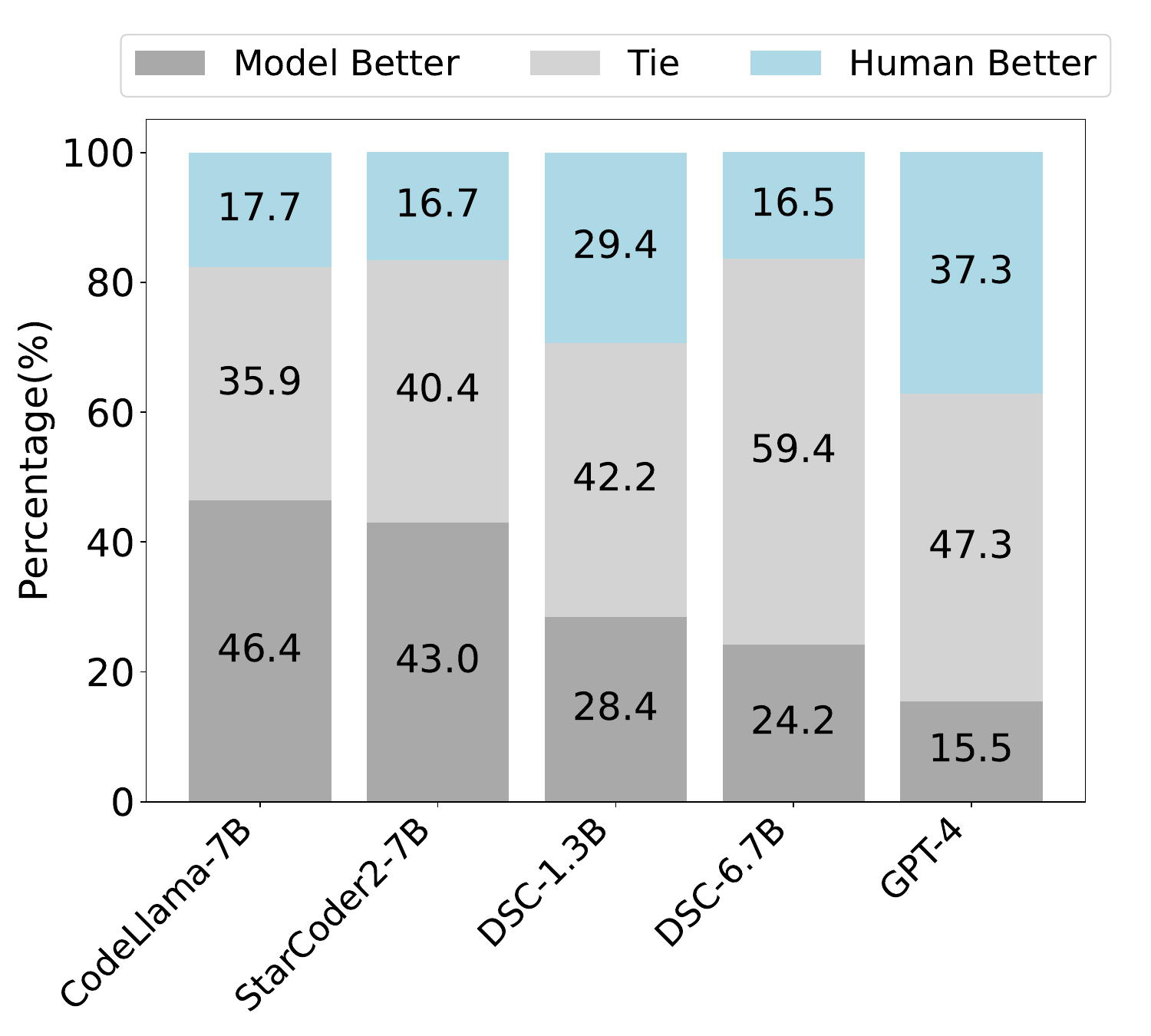}
}\hspace{-5pt}
\subfigure[Robustness]{
    \label{fig:rq3-robustness}
    \includegraphics[width=0.33\linewidth]{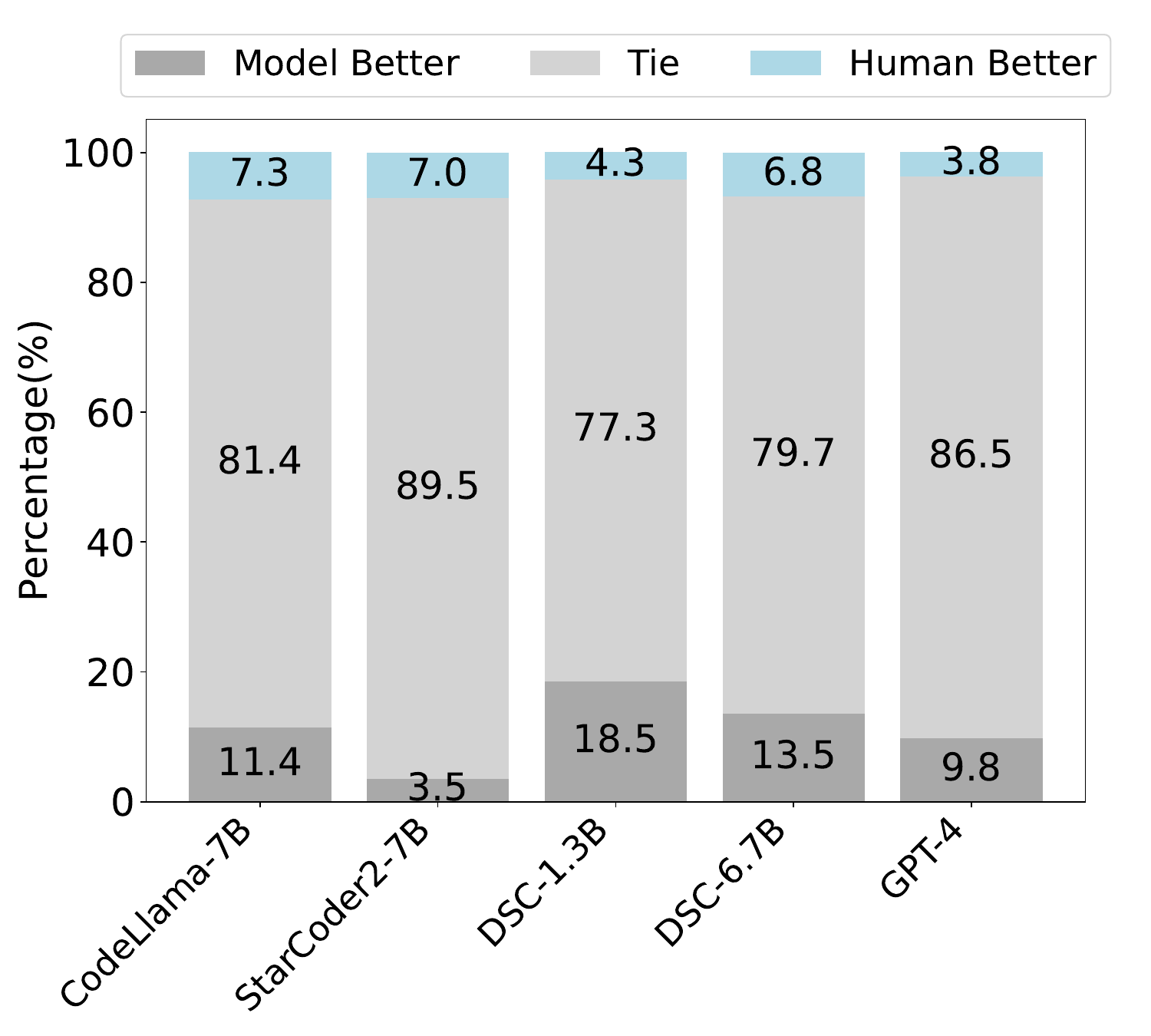}
}\hspace{-25pt}
% \vspace{-10pt}
\figmargin
\figmargin
\caption{Score Distribution across Readability, Conciseness, and Robustness.}
\label{fig:rq3}
\end{figure}

The annotation is conducted independently by two annotators. Both annotators have many years of programming experience and possess a certain level of understanding regarding research on code style. 
The two annotators independently score the code samples, resulting in two sets of annotations. Cohen's Kappa~\cite{cohen1960coefficient} is a statistical measure used to assess the consistency between two annotators for a classification task. We calculated the Cohen's Kappa values for the two annotations on readability, conciseness, and robustness, which are 0.63, 0.79, and 0.68, respectively. However, the discrepancies that arose in the annotations were ultimately resolved through discussions to reach a consensus.
Figure~\ref{fig:rq3} show the proportion of code samples that received different scores (model better, tie and human better) on the three aspects for each model. Overall, the code samples generated by the LLMs is comparable to that written by human programmers in the three aspects. On average, the code generated by the five models is comparable to or even superior to the code written by programmers in 89.5\%, 76.5\%, and 94.2\% of cases in terms of readability, conciseness, and robustness, respectively. The following is a comparative analysis of the readability, conciseness, and robustness of the code samples generated by the five LLMs. From the perspective of readability, the code samples generated by \deepbig{} have the highest readability, while those generated by \starcoder{} have the lowest. In terms of conciseness, the conciseness of code samples generated by \codellama{}, \starcoder{}, and \deepbig{} is comparable, while \gptfour{} generates less concise code. 
Figure~\ref{fig:conciseness} presents an example that the conciseness of a code sample generated by \gptfour{} is inferior to that of ground truth written by human programmers. Note that conciseness and readability are often trade-offs. As shown in Figure~\ref{fig:conciseness}, \gptfour{} enhances readability by splitting one statement into two lines. All five studied LLMs demonstrate relatively high robustness. This suggests that the models might have learned more robust coding styles from their training data, such as more rigorous input parameter checks, which human programmers might omit due to oversight or to avoid excessive complexity.

\begin{center}
    % \begin{myboxb}[]{RQ3 Summary} %ab
    \begin{myboxc} \textbf{RQ3 Summary: } %cd
  % Overall, code generated by Code LLMs is comparable to or even better than human-written code in terms of readability, conciseness, and robustness. 
  % Among the studied models, \deepbig{} produces the most readable code, while \codellama{} and \deepsmall{} lags in readability and conciseness, respectively.
  Overall, the code samples generated by the five LLMs is comparable to human-written code in terms of readability, conciseness, and robustness. 
  Among the studied models,  \deepbig{} produces the most readable code, while code samples generated by \gptfour{} has the lowest simplicity but the highest robustness.
    \end{myboxc} %cd
    % \end{myboxd} %cd
    % \end{myboxb} %ab
\end{center}

\subsection{RQ4: Style Improvement by Prompting Techniques}
\label{sec:rq3:design}

\begin{figure}[t]
\centering
%\vspace{-5mm}
\includegraphics[width=0.9\columnwidth]{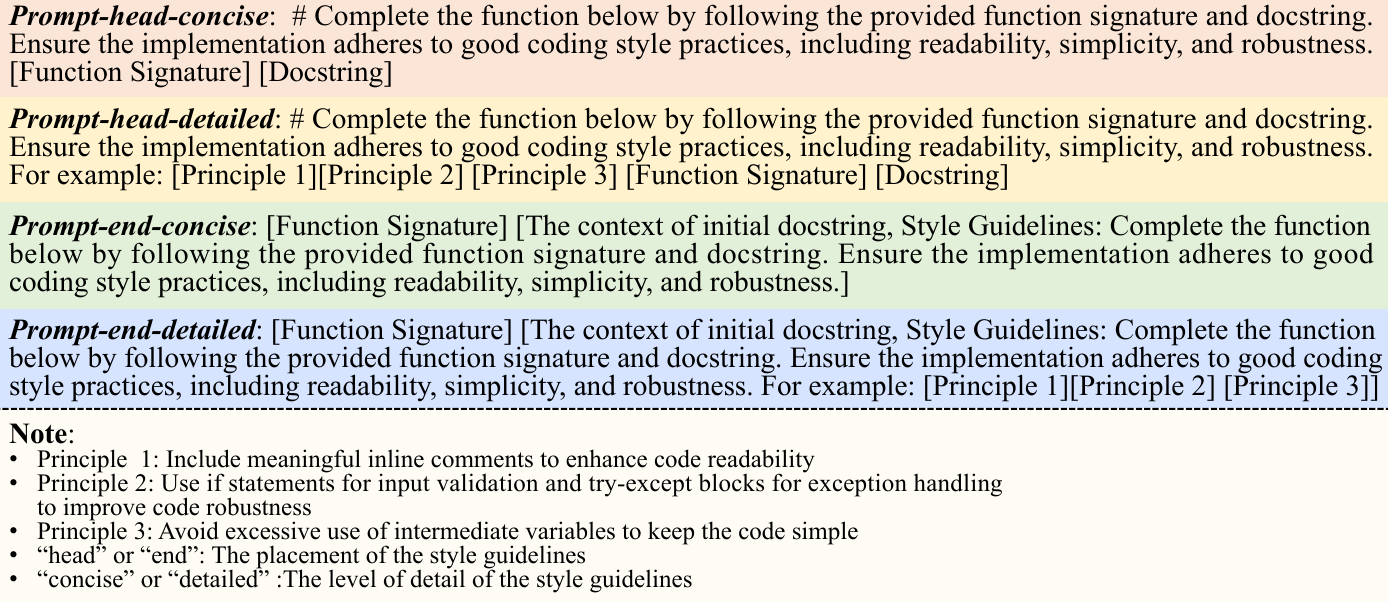}
\figmargin
\caption{Four Enhanced Prompts in RQ4 Study.} 
\label{fig:prompts}
\figmargin
\vspace{10pt}
\end{figure}

In this RQ, we investigate whether prompting techniques can improve the coding style of LLMs. We conduct experiments with \deepbig{} on 20 sampled Python tasks from CoderEval. 
% We choose \deepbig{} to conduct the experiment with type a because it achieves the best functional correctness in generating functions among the four models.
These tasks are randomly selected from those that \deepbig{} can complete, meaning \deepbig{} can generate code samples that pass all corresponding test cases.
We design four types of enhanced prompts for this study (refer to Figure ~\ref{fig:prompts}), aiming to instruct the model to generate code with better coding style using explicit style guidelines. The design of these prompts investigates the impact of the placement and detail level of style guidelines.
In prompt names, ``-head'' or ``-end'' specifies whether the style guidelines are placed before the function signature and docstring, similar to a directive, or appended at the end of the original docstring, simulating a normal docstring style. ``-concise'' and ``-detailed'' indicate the level of detail in the style guidelines. The detailed version includes three specific principles related to code readability, conciseness, and robustness, in addition to the concise information.

Among the selected tasks, \deepbig{} generates 134 functionally correct code samples using the basic prompt, i.e., the original function signature and docstring as input. Then, for each type of enhanced prompt, \deepbig{} generates 10 code samples for the 20 selected tasks, resulting in 115, 137, 75, and 78 functionally correct code samples for each of the four enhanced prompts, respectively. 
The accuracy for the four enhanced prompts is 57.5\%, 68.5\%, 37.5\%, and 39.0\%, respectively, compared to the 67.0\% accuracy of the basic prompt.
Except for prompt-head-detailed, the enhanced prompts result in lower accuracy compared to the basic prompt, suggesting that using more complex prompts may lead to a decrease in the functional correctness of the generated code.

% height=0.7\linewidth
\begin{figure}[t]
\centering
\hspace{-10pt}
\subfigure[Readability]{
    \label{fig:rq4-readability}
    \includegraphics[width=0.33\linewidth]{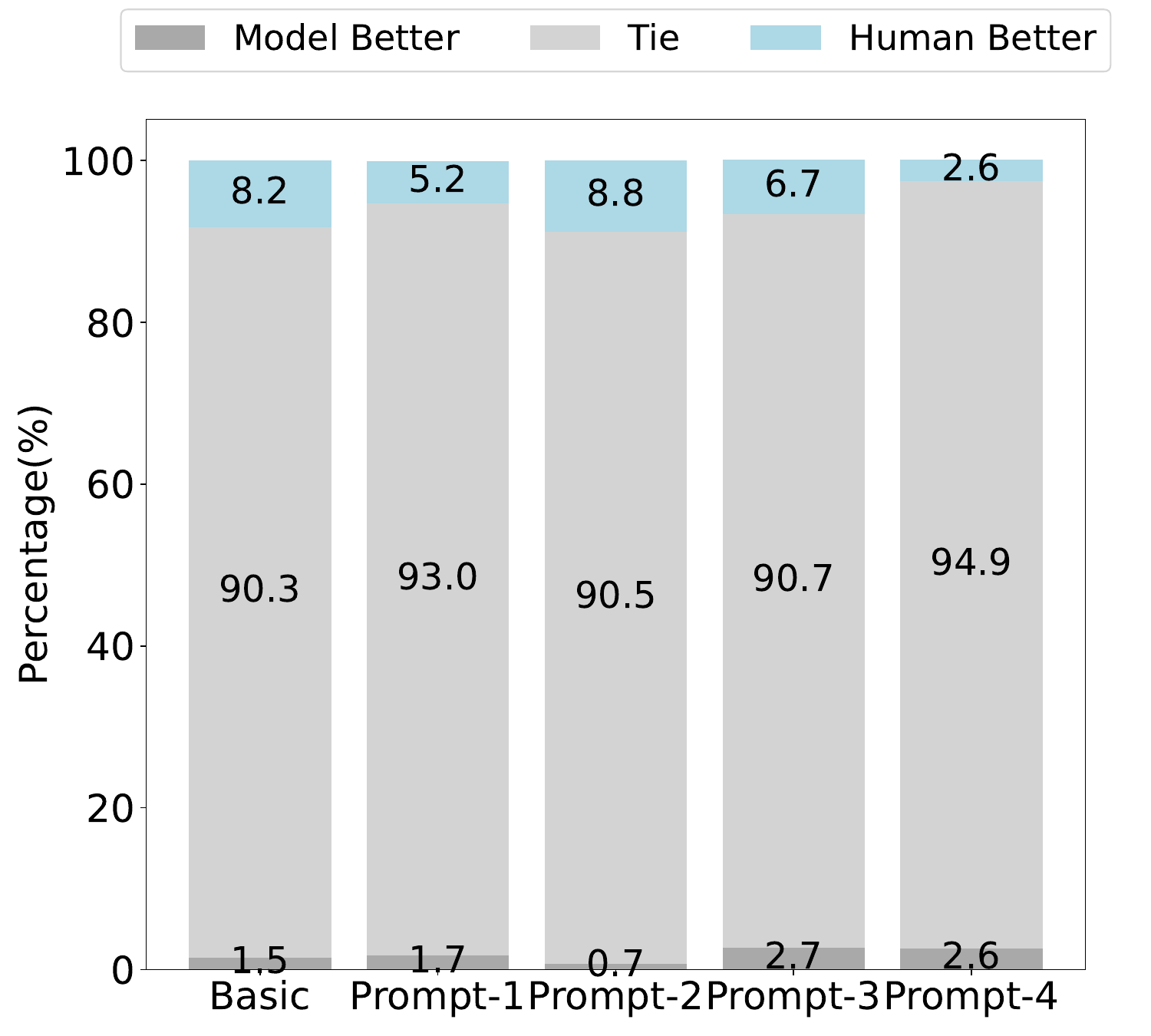}
    \figmargin
}\hspace{-8pt}
\subfigure[Conciseness]{
    \label{fig:rq4-conciseness}
    \includegraphics[width=0.33\linewidth]{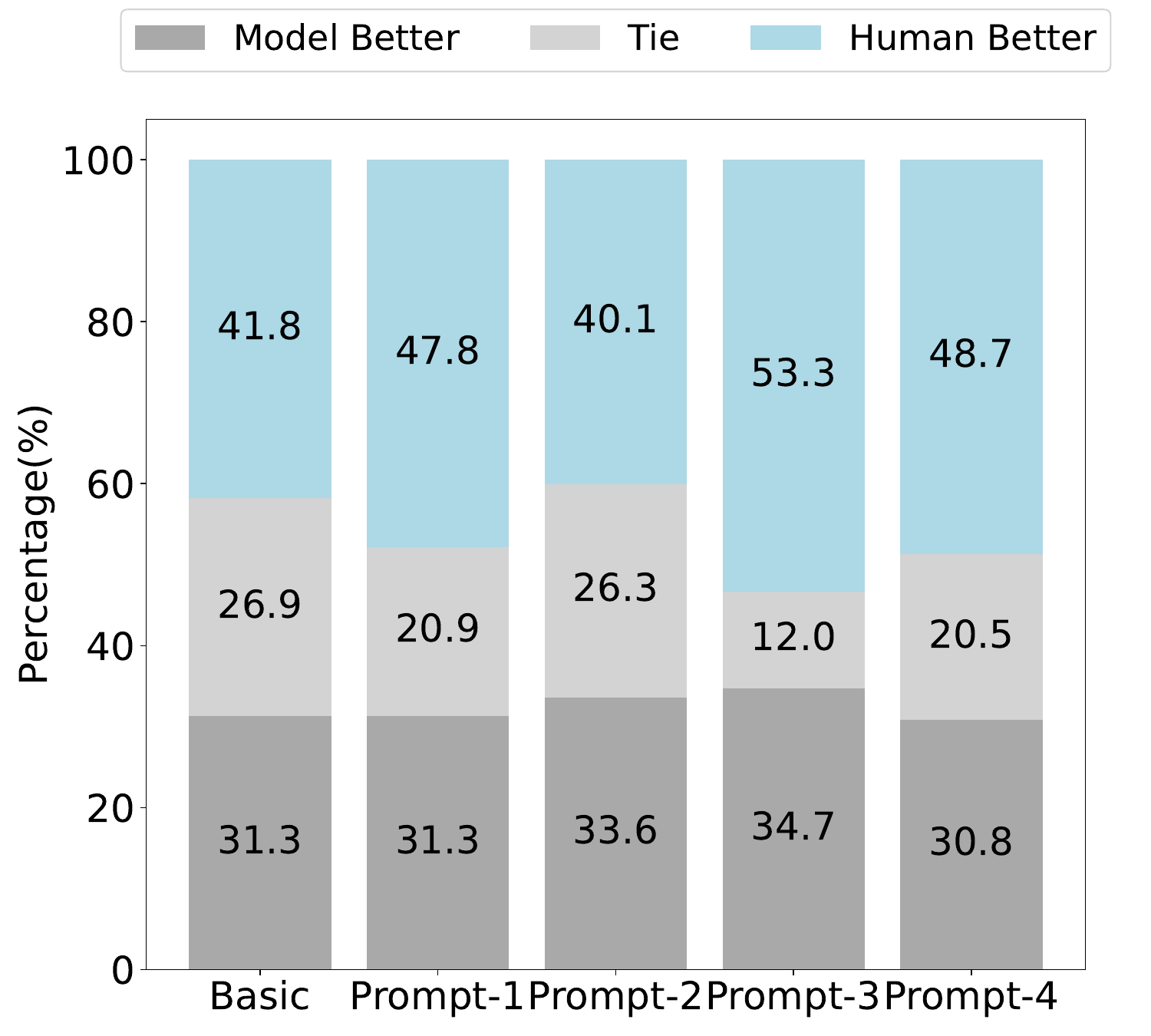}
    \figmargin
}\hspace{-8pt}
\subfigure[Robustness]{
    \label{fig:rq4-robustness}
    \includegraphics[width=0.33\linewidth]
    {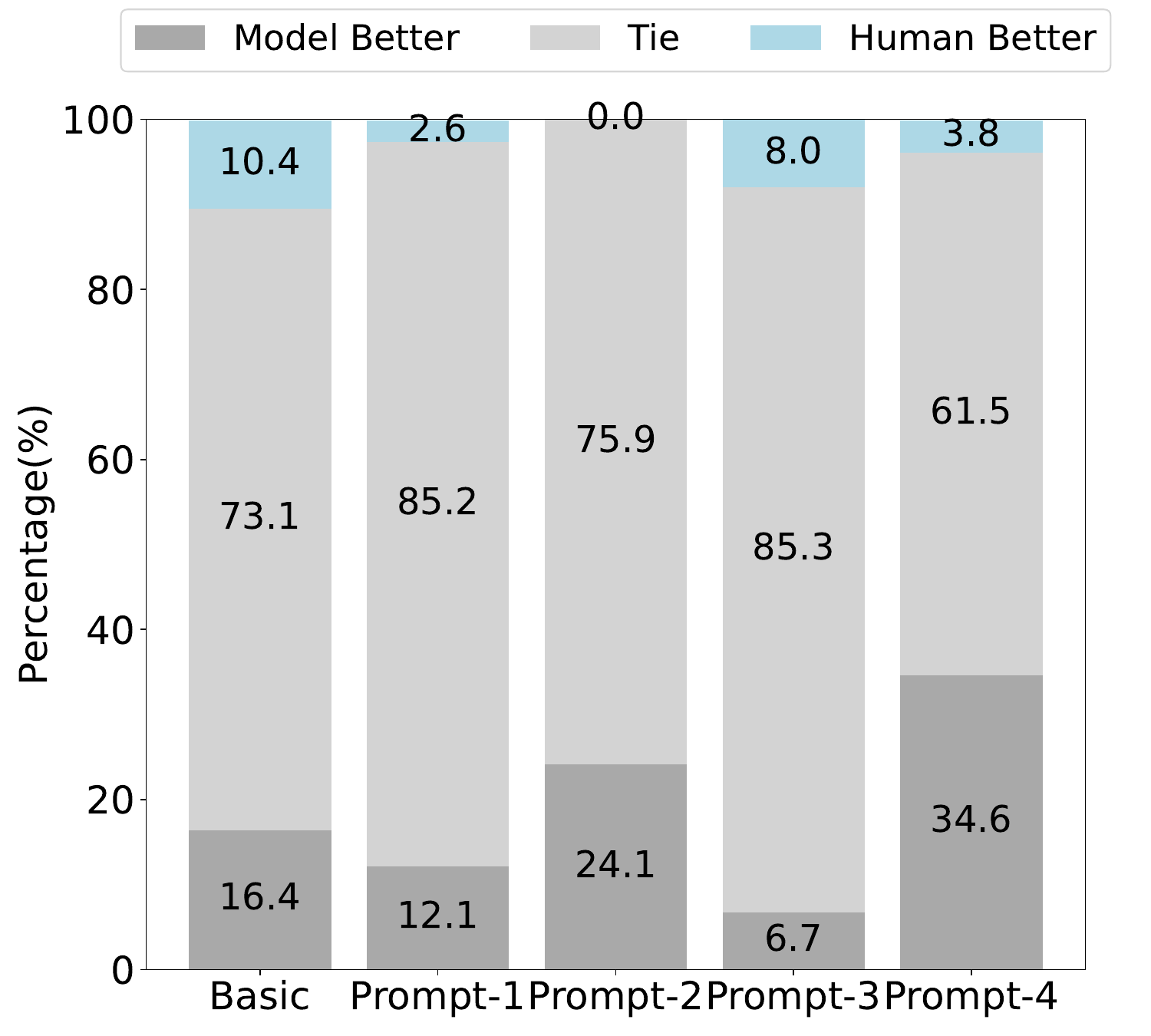}
    \figmargin
}
\hspace{-25pt}
\vspace{-5pt}
\figmargin
\caption{Score Distribution across Readability, Conciseness, and Robustness by Different Prompts. P-1, P-2, P-3, P-4 Stand for Prompt-head-concise, Prompt-head-detailed, Prompt-end-concise, and Prompt-end-detailed, Respectively.}
\label{fig:rq4}
\end{figure}

% According to the scoring principles outlined in Section~\ref{sec:rq3}, the annotators metioned in  evaluated the code samples generated using the basic prompt and four enhanced prompts for readability, conciseness, and robustness. 
The two annotators mentioned in Section~\ref{sec:rq3}
referred to the scoring principles mentioned in Section~\ref{sec:rq3}
to score the code samples generated using the basic prompt and the four enhanced prompts for readability, conciseness, and robustness. 
The Cohen's Kappa for the two sets of annotations on readability, conciseness, and robustness were 0.65, 0.81, and 0.74, respectively. Despite the discrepancies in the annotations, the annotators were eventually resolved through discussions, leading to a consensus.
%% need revise 
The G-test~\cite{rohlf1981biometry} is a statistical method used to assess the association between categorical variables. In this study, we applied the G-test to determine whether the probability distributions of scores for the basic prompt and the four enhanced prompts show significant differences across the three aspects of readability, conciseness, and robustness. The results indicated that the p-values for all three aspects were below 0.05, suggesting a statistically significant difference in the score distributions among the five prompts across these three aspects.

The results are depicted in Figure~\ref{fig:rq4}. 
Among the enhanced prompts, Prompt-head-concise, Prompt-end-concise, and Prompt-end-detailed slightly improve the readability of the code samples generated by \deepbig{}. However, as shown in Figure~\ref{fig:rq4-conciseness}, only Prompt-head-detailed enhances the conciseness of \deepbig{}'s code samples. This is because there's often a trade-off between readability and conciseness, where improving one may compromise the other. Additionally, as seen in Figure~~\ref{fig:rq4-robustness}, all four enhanced prompts contribute to some extent to the improved robustness of \deepbig{}'s code samples.
In conclusion:
(i) Incorporating style-guiding information into prompts may lead to decreased accuracy in generated code, as observed in our evaluation.
(ii) Relying solely on prompt engineering may not fully resolve issues related to code style. Additional strategies or refinements may be necessary.

\begin{center}
    \begin{myboxc} \textbf{RQ4 Summary: } %cd
    Certain types of prompts can slightly improve the readability and robustness of generated code, but only one type enhances conciseness. There is a trade-off between readability and conciseness, indicating that while prompt engineering can help, it is not sufficient to fully address issues related to coding style. Including guidance in prompts may also decrease the accuracy of generated code.
    \end{myboxc} %cd
    % \end{myboxd} %cd
    % \end{myboxb} %ab
\end{center}

\subsection{Case Studies}
In the code samples we observed, we categorized and analyzed cases where the code samples exhibited inconsistent coding styles compared to the ground truth. We identified the following common scenarios.

\begin{figure}[t]
\centering
%\vspace{-5mm}
\includegraphics[width=\columnwidth]{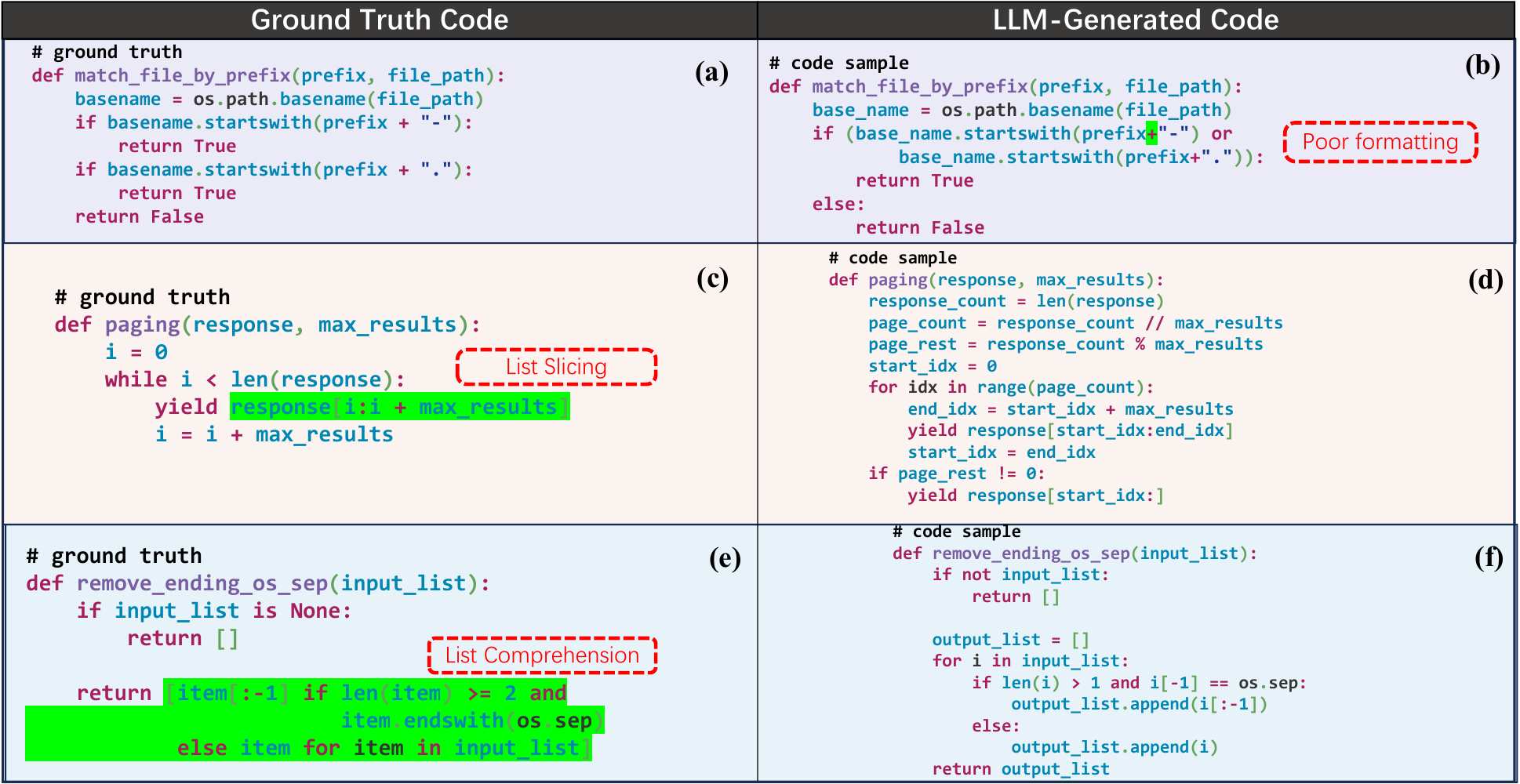}
\figmargin
\vspace{-10pt}
\caption{Examples of (i) Poor formatting characteristics; (ii) Unfamiliar with basic Python features; (iii) Rare use of advanced syntax features. On the left side of the image, (a), (c), and (e) are the ground truth for each example, while on the right side, (b), (d), and (f) are the code samples generated by LLMs.} 
\label{fig:casestudy}
\figmargin
% \vspace{-5pt}
\end{figure}

% \begin{mdframed}[backgroundcolor=yellow!25,hidealllines=true,innerleftmargin=0pt,innerrightmargin=0pt,innertopmargin=0pt,innerbottommargin=0pt, skipabove=0pt, skipbelow=0pt]
% (1) Remove the example of the deprecated API. (2) Expanded 4.5.
% \end{mdframed}

% \textbf{Using deprecated APIs.}
% In Figure~\ref{fig:casestudy} (b), the code sample generated by \codellama{} uses the \texttt{getchildren()} method, which was deprecated in Python 3.2 and removed in Python 3.9. This might be due to \codellama{} being trained on a corpus that includes Python code from different versions, leading to unawareness that certain APIs are outdated. Including deprecated APIs in code generated by large models is considered bad coding style, as this code will produce errors when run on newer Python versions.

\textbf{Poor formatting characteristics.}
LLMs may generate code with poor formatting characteristics, such as improper use of spaces. Figure~\ref{fig:casestudy} (b) is a function generated by \gptfour{}. In the expression \textit{``base\_name.startswith(prefix+"-")''}, the ``+'' operator is not surrounded by spaces, which is not a good practice in the use of spaces. The code generated by the other four models also contains improper use of spaces, such as \deepbig{} generating the code \textit{``yield response[start: end]''}. However, the PEP-8 compliant version should be written as \textit{``yield response[start:end]''}.

\textbf{Unfamiliar with basic Python features.} %Unfamiliar with the basic features of Python.} 
LLMs might not be very familiar with some basic syntax features, which results in generating more complex code. For example, in Figure~\ref{fig:casestudy} (d), \deepbig{} might not understand list slicing operations well, so it generated more complex code to avoid out-of-bounds indexing. 
Assuming the list has a length of 4, using \texttt{list[3:5]} in Python will not result in an error. Instead, it will return elements from index 3 to the end of the list. However, in the corresponding ground truth of the code sample (Figure~\ref{fig:casestudy} (c)), the code logic is clear and concise.

\textbf{Rare use of advanced syntax features.}
Compared to code written by human programmers, code generated by LLMs often does not use advanced syntax features of the Python language, such as Pythonic idioms. As shown in Figure~\ref{fig:casestudy} (e), the ground truth uses list comprehension to build a list, while the code sample in Figure~\ref{fig:casestudy} (f), uses a more conventional method to build the list. It first constructs an empty list and then uses the \texttt{append()} method to add elements to the empty list. Compared to the ground truth, the simplicity of the code sample is inferior.

\section{Threats to Validity}
We have identified the following threats to our study.

\textbf{Data Quality.}
One potential threat to validity is the quality of the raw data used for our empirical study. To ensure the quality of the data for open coding, we applied multiple strategies: comprehensive unit testing to validate the functionality of the generated code samples, manual filtering to remove any that did not meet our criteria for functional correctness and implementation conciseness, and selecting tasks from the popular benchmark CoderEval, ensuring their high quality and relevance.

\textbf{LLM Utilization.}
Another potential threat is the utilization (e.g., source, parameter settings) of the LLMs used in our study. We carefully used the official release versions of each model to avoid any potential issues with unofficial or modified versions, followed the guidelines provided by the model developers to ensure proper implementation and usage, and conducted repeated tests to verify the performance and consistency of the models' outputs. To ensure a fair comparison, we used the same prompt structure and generation parameters for each model, standardizing the experimental setup across different models.

\textbf{Taxonomy Reliability and Completeness.}
The reliability and completeness of the inconsistency types identified %during our study 
pose another potential threat. We employed the open coding methodology to systematically identify and categorize inconsistency types, adhered to established open coding practices to ensure thoroughness and accuracy, and ensured that our taxonomy was stable by iteratively refining the inconsistency types until no new categories emerged. 
We involved multiple annotators to score these metrics, and they discussed their ratings to reach a consensus, reducing individual biases and ensuring more objective assessments. To further bolster the credibility of our findings, we have made all our data publicly available, allowing others to verify our results and methodology, thus enhancing the robustness of our conclusions.

\vspace{-10pt}
\section{Conclusion}
Many studies have focused on improving the functional correctness of LLM-based code generation. However, the code style of LLMs—an important aspect of code quality that extends beyond functional correctness—remains under-explored. To fill this gap, this paper makes the first attempt to investigate the coding style differences between LLMs and human developers through an empirical study. Specifically, we compare the code generation results of five mainstream LLMs with ground truth on the CoderEval benchmark.

We present a comprehensive taxonomy of coding style inconsistencies between LLMs and human developers, identifying 24 inconsistency types across five dimensions. We adopt the definition of code style from previous work, but our taxonomy offers a more detailed classification of code style. Our analysis reveals clear coding style differences between the studied LLMs and human developers, particularly in statements/expressions and formatting, while showing similar coding styles among the studied LLMs. We further discuss potential causes of these style inconsistencies and explore ways to improve coding style discrepancies through prompt engineering, providing a foundation for future research in this area.

\section{Data Availability}
All the data used in this study is provided in the replication package \cite{replication_package}.

\section{Acknowledgments}
The work is supported by CCF-Huawei Populus Grove Fund CCFHuaweiSE202403, the Natural Science Foundation of Guangdong Province under Grant 2024A1515010255, and National Natural Science Foundation of China under Grant No. 62402113.